\documentclass[nonacm, sigconf]{acmart}

\usepackage{graphicx} %
\usepackage{cleveref}
\usepackage{subcaption}
\usepackage{rotating}

\crefformat{section}{Section~#2#1#3}
\crefformat{subsection}{Section~#2#1#3}
\crefformat{equation}{Eq.~#2#1#3}
\crefformat{figure}{Fig.~#2#1#3}

\usepackage{fontawesome}
\newcommand{\videotime}[2]{\faPlayCircle\,#1m#2s}

\newcommand{\todo}[1]{}
\newcommand{\ryusuke}[1]{}
\newcommand{\toshiya}[1]{}
\newcommand{\christopher}[1]{}

\newcommand{\dimension}{d}
\newcommand{\pres}{p}
\newcommand{\ppres}{\overline{p}} %
\newcommand{\vel}{\mathbf{u}}
\newcommand{\dens}{\rho}
\newcommand{\wobdens}{\mu}
\newcommand{\visc}{\nu}
\newcommand{\Rd}{{\mathbb{R}^\dimension}}
\newcommand{\domain}{\Omega}
\newcommand{\sdomain}{{\domain^s}}
\newcommand{\boundary}{{\partial\Omega}}
\newcommand{\sboundary}{{\partial\sdomain}}
\newcommand{\laplace}{\nabla^2}
\newcommand{\grad}{\nabla}

\newcommand{\diverg}{\nabla \cdot}
\renewcommand{\vec}{\mathbf}
\newcommand{\vecx}{\vec{x}}
\newcommand{\vecy}{\vec{y}}

\newcommand{\vecr}{\vec{r}}
\newcommand{\vecn}{\vec{n}}

\newcommand{\fund}{G}
\newcommand{\dt}{\Delta t}
\newcommand{\transpose}{^\intercal}
\newcommand{\identity}{\mathbf{I}}
\newcommand{\dVy}{\,\mathrm{dV}(\vecy)}
\newcommand{\dAy}{\,\mathrm{dA}(\vecy)}

\newcommand{\velsol}{\vel_s}
\newcommand{\dGdnx}{{\frac{\partial \fund}{\partial \vecn_\vecx}}}

\newcommand{\acc}{\vec{f}}
\newcommand{\gravity}{\vec{g}}
\newcommand{\temperature}{T}
\newcommand{\concentration}{s}
\newcommand{\estimator}[1]{\langle{#1}\rangle}
\newcommand{\diffusionfund}{Z}
\newcommand{\dZdny}{{\frac{\partial \diffusionfund}{\partial \vecn_\vecy}}}
\newcommand{\diffusionwobdens}{\phi}
\newcommand{\diffusionsol}{w}
\newcommand{\heaviside}{\Theta}
\newcommand{\dtau}{\,\mathrm{d}\tau}
\newcommand{\pdf}{P}

\defcitealias{RiouxLavoie2022:Fluids}{Rioux-Lavoie and Sugimoto et~al.}
\defcitealias{Sawhney:2022:GFMC}{Sawhney and Seyb et~al.}
\defcitealias{sawhney2023walk}{Sawhney and Miller et~al.}

\newcommand{\mycite}[1]{[\citetalias{#1}~\citeyear{#1}]}
\newcommand{\mycitet}[1]{\citetalias{#1}~[\citeyear{#1}]}

\title{Velocity-Based Monte Carlo Fluids}

\author{Ryusuke Sugimoto}
\orcid{0000-0001-5894-0423}
\affiliation{
    \institution{University of Waterloo}
    \city{Waterloo}
    \state{Ontario}
    \country{Canada}
}
\email{rsugimot@uwaterloo.ca}
\author{Christopher Batty}
\orcid{0000-0003-3830-7772}
\affiliation{
    \institution{University of Waterloo}
    \city{Waterloo}
    \state{Ontario}
    \country{Canada}
}
\email{christopher.batty@uwaterloo.ca}
\author{Toshiya Hachisuka}
\orcid{0000-0003-0284-3776}
\affiliation{
    \institution{University of Waterloo}
    \city{Waterloo}
    \state{Ontario}
    \country{Canada}
}
\email{toshiya.hachisuka@uwaterloo.ca}

\begin{document}



\begin{abstract}
We present a velocity-based Monte Carlo fluid solver that overcomes the limitations of its existing vorticity-based counterpart. 
Because the velocity-based formulation is more commonly used in graphics, our Monte Carlo solver can be readily extended with various techniques from the fluid simulation literature.
We derive our method by solving the Navier-Stokes equations via operator splitting and designing a pointwise Monte Carlo estimator for each substep.
We reformulate the projection and diffusion steps as integration problems based on the recently introduced walk-on-boundary technique~\cite{Sugimoto2023}.
We transform the volume integral arising from the source term of the pressure Poisson equation into a form more amenable to practical numerical evaluation.
Our resulting velocity-based formulation allows for the proper simulation of scenes that the prior vorticity-based Monte Carlo method~\mycite{RiouxLavoie2022:Fluids} either simulates incorrectly or cannot support.
We demonstrate that our method can easily incorporate advancements drawn from conventional non-Monte Carlo methods by showing how one can straightforwardly add buoyancy effects, divergence control capabilities, and numerical dissipation reduction methods, such as advection-reflection and PIC/FLIP methods.\looseness=-1
\end{abstract}

\begin{teaserfigure}
    \centering
    \begin{minipage}[b]{0.52\linewidth}
        \includegraphics[trim={1cm 1cm 1cm 1.4cm}, clip, width=0.19\linewidth]{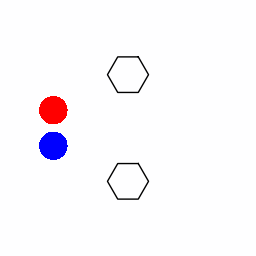}
        \includegraphics[trim={1cm 1cm 1cm 1.4cm}, clip, width=0.19\linewidth]{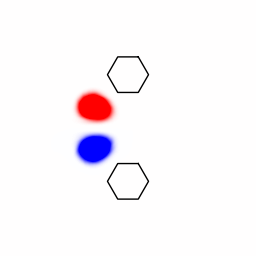}
        \includegraphics[trim={1cm 1cm 1cm 1.4cm}, clip, width=0.19\linewidth]{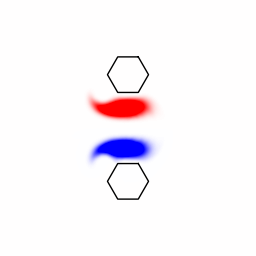}
        \includegraphics[trim={1cm 1cm 1cm 1.4cm}, clip, width=0.19\linewidth]{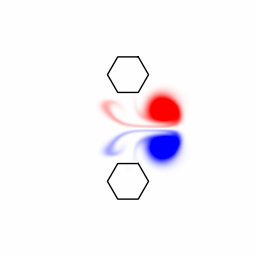}
        \includegraphics[trim={1cm 1cm 1cm 1.4cm}, clip, width=0.19\linewidth]{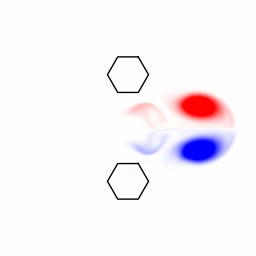}\vspace{-0.2cm}\\
        \small\textbf{\textsf{Correct Physics with Velocity-Based Monte Carlo Method (Ours)}}\\
        \includegraphics[trim={1cm 1cm 1cm 0.7cm}, clip, width=0.19\linewidth]{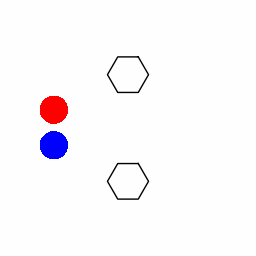}
        \includegraphics[trim={1cm 1cm 1cm 0.7cm}, clip, width=0.19\linewidth]{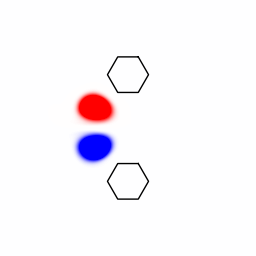}
        \includegraphics[trim={1cm 1cm 1cm 0.7cm}, clip, width=0.19\linewidth]{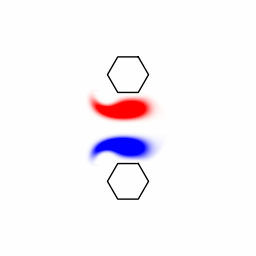}
        \includegraphics[trim={1cm 1cm 1cm 0.7cm}, clip, width=0.19\linewidth]{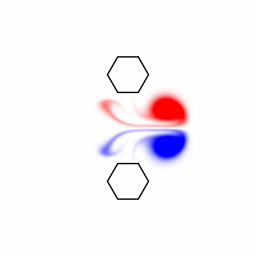}
        \includegraphics[trim={1cm 1cm 1cm 0.7cm}, clip, width=0.19\linewidth]{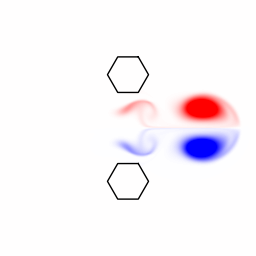}\vspace{-0.2cm}\\
        \small\textbf{\textsf{Correct Physics with Conventional Velocity-Based Method\\\cite{Batty2007}}}\\
        \includegraphics[trim={1cm 0.5cm 1cm 0.7cm}, clip, width=0.19\linewidth]{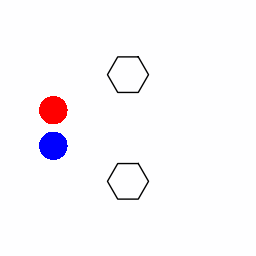}
        \includegraphics[trim={1cm 0.5cm 1cm 0.7cm}, clip, width=0.19\linewidth]{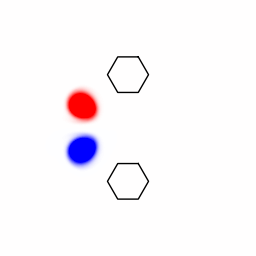}
        \includegraphics[trim={1cm 0.5cm 1cm 0.7cm}, clip, width=0.19\linewidth]{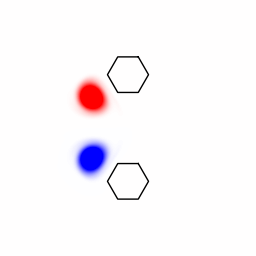}
        \includegraphics[trim={1cm 0.5cm 1cm 0.7cm}, clip, width=0.19\linewidth]{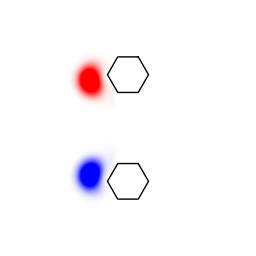}
        \includegraphics[trim={1cm 0.5cm 1cm 0.7cm}, clip, width=0.19\linewidth]{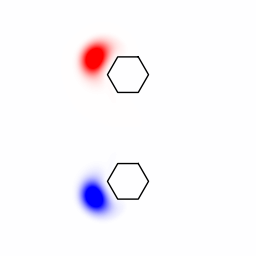}\vspace{-0.3cm}\\
        \small\textbf{\textsf{Incorrect Physics with Vorticity-Based Monte Carlo Method\\\mycite{RiouxLavoie2022:Fluids}}} \hfill\small\textsf{\videotime{0}{56}}\;\;\;
    \end{minipage}
    \begin{minipage}[b]{0.47\linewidth}
        \includegraphics[trim={0, 0.5cm, 0, 1cm}, clip, width=\linewidth]{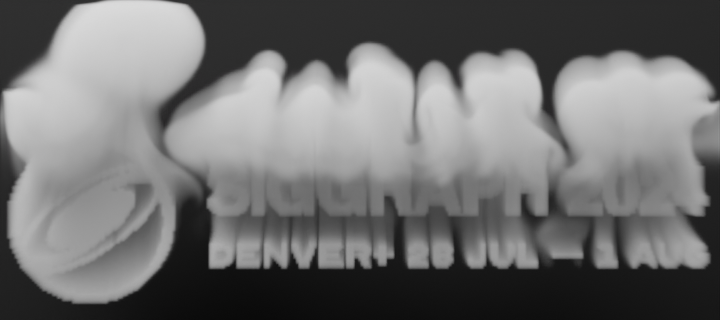}
        \includegraphics[trim={0, 3cm, 0, 0}, clip, width=0.497\linewidth]{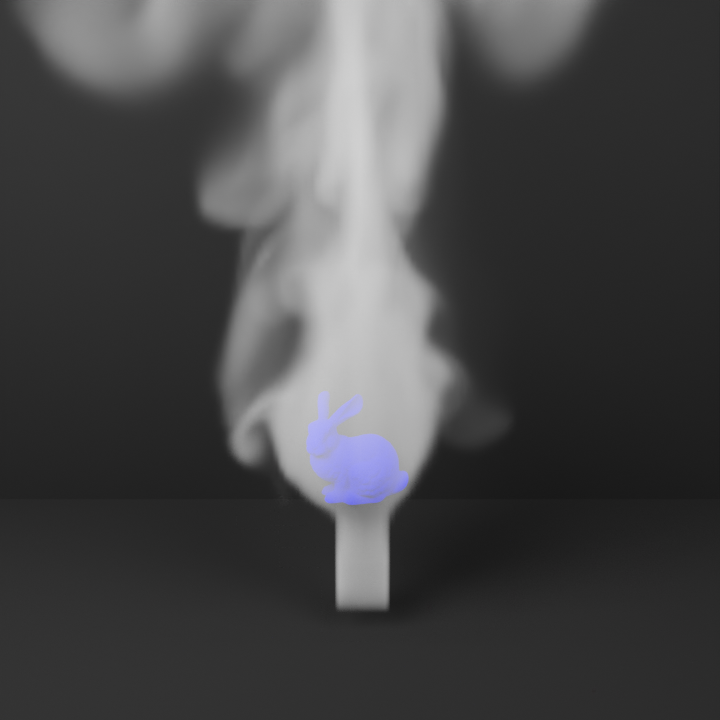}
        \includegraphics[trim={0, 3cm, 0, 0}, clip, width=0.497\linewidth]{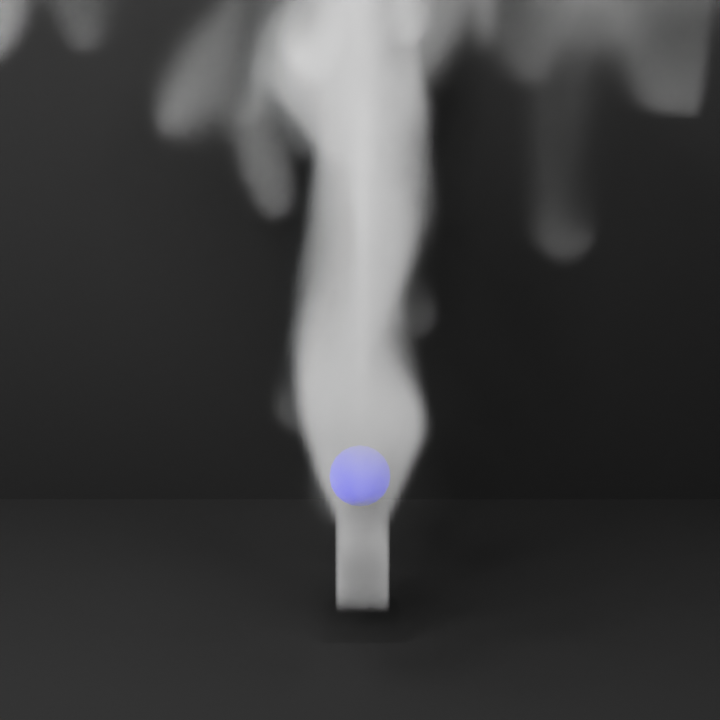}
        \vspace*{-2.8em}
        \begin{center}
        \hfill\small\textsf{\textcolor{white}{\videotime{0}{15}}\;}
        \end{center}   
    \end{minipage}
    \caption{Our velocity-based Monte Carlo fluid solver can naturally handle scenes for which the existing vorticity-based Monte Carlo method~\mycite{RiouxLavoie2022:Fluids} yields incorrect results. Our solver allows the red and blue smoke densities associated with a pair of vortices to flow between two obstacles (left, top), similar to the conventional non-Monte Carlo velocity-based method~\cite{Batty2007} (left, middle), while the vorticity-based method produces an incorrect result, in which smoke deviates around the outside of the obstacles (left, bottom). Our solver also readily supports commonly available features in traditional velocity-based Eulerian grid fluid solvers, such as buoyancy effects: we simulate a smoke plume rising from a complex-shaped source (right, top) and past a sphere-shaped or bunny-shaped obstacle (right, bottom), where the motion is driven solely by buoyancy forces.}
    \label{fig:teaser}
\end{teaserfigure}

\maketitle

\section{Introduction}
Researchers in computer graphics have recently revisited Monte Carlo PDE solvers~\cite{Muller:1956:Some,sabelfeld1982vector} due to their attractive properties. 
\citet{Sawhney:2020:MCG} showed that geometry processing tasks can reap benefits from Monte Carlo solvers, such as flexibility with respect to geometric boundary representations, robustness to noise in the input geometry, support for pointwise estimation of the solution, and trivial parallelization.
Monte Carlo methods have already been used in light transport simulation for decades since the work of \citet{Kajiya:1986:Rendering} due to these same properties.\looseness=-1

Looking beyond light transport and geometry processing, \mycitet{RiouxLavoie2022:Fluids} developed the first Monte Carlo method to perform smoke simulation based on the Navier-Stokes equations. 
They discussed theoretical and practical aspects of Monte Carlo-based fluid simulation; most advantages in the geometry processing context carry over directly to fluid simulation applications. 
However, their method relies on a vorticity-based formulation, whereas velocity-based formulations are more common in fluid animation.

Since the introduction of velocity-based Eulerian grid fluid simulation techniques to graphics~\cite{FOSTER1996:Realistic, Stam1999}, researchers have improved on many aspects of the velocity-based formulation, adding features such as 
buoyancy modeling for smoke~\cite{Fedkiw2001, Foster1997} and velocity divergence control for expansion/contraction and artistic effects~\cite{Feldman2003:Animating}, as well as 
PIC/FLIP solvers \cite{zhu2005animating, Harlow1962:PIC, BRACKBILL1986:FLIP} and advection-reflection solvers~\cite{Zehnder2018, Narain2019} for reducing numerical dissipation.
These and many other improvements cannot be straightforwardly incorporated into the Monte Carlo fluid solver of \mycitet{RiouxLavoie2022:Fluids} because of its vorticity-based formulation. Furthermore, \citet{Yin2023} recently showed that existing vorticity-based methods can fail to simulate harmonic velocity fields, leading to incorrect results when the fluid domain is not simply connected (e.g., is disjoint or has holes). Conveniently, however, velocity-based methods can capture these physics without any change.

We therefore introduce a %
Monte Carlo fluid solver that relies on a velocity-based formulation. 
Similarly to the vorticity-based method \mycite{RiouxLavoie2022:Fluids}, our method does not require the boundaries of solid objects to be explicitly discretized; for example, there is no need for a cut-cell method~\cite{Batty2007} or a conforming mesh~\cite{Feldman2005} as in traditional solvers, and the only requirement on the geometry is that it support ray intersection queries.
We use the recently introduced walk-on-boundary method~\cite{Sugimoto2023}, which is a Monte Carlo ray tracing solver for PDEs, to perform pressure projection in our solver, and thus, our solver can be trivially accelerated by GPU ray tracing.
Our method can additionally incorporate all of the aforementioned techniques developed for velocity-based simulations: buoyancy, divergence control, PIC/FLIP, and advection-reflection. 

To solve the Navier-Stokes equations, we apply operator splitting~\cite{Stam1999}, and the core of our solver depends on reformulating the projection and diffusion steps as integration problems. In the absence of obstacles, we derive a non-recursive integral for each step and apply Monte Carlo integration. The projection step requires a careful evaluation of the volume integral term arising from the Poisson equation's source term, and we describe in detail how to transform it into a form that is more amenable to numerical evaluation. Domains with boundaries require solving a boundary integral equation, which we treat using the walk-on-boundary method~\cite{Sugimoto2023}. The diffusion equation for the diffusion step is a time-dependent equation, and we apply a \emph{time-dependent} walk-on-boundary method~\cite{SabelfeldSimonov1994} for problems with boundaries; we are the first to apply it to a computer graphics application. After constructing Monte Carlo solvers for each substep, we combine them to design a Navier-Stokes solver with flexible options for how we store the intermediate velocity field in a discretized cache.
To summarize, our contributions include:%
\begin{itemize}
    \item a velocity-based Monte Carlo fluid simulator that uses operator splitting,
    \item an integral formulation for the projection step with careful Poisson source term handling,
    \item extension of the projection step to problems with boundaries using the walk-on-boundary method,
    \item application of the diffusion walk-on-boundary method for the diffusion step, 
    \item adaptations of more advanced velocity-based solver techniques to our Monte Carlo formulation.
\end{itemize}

\section{Method}
Our method aims to numerically solve for a velocity field that satisfies the incompressible Navier-Stokes equations with constant density and viscosity:
\begin{equation}\label{eq:ns}
\begin{gathered}
        \frac{\partial \vel}{\partial t} = - (\vel \cdot \grad)  \vel - \frac{1}{\dens}\grad \pres + \visc  \laplace \vel + \acc,\\
    \diverg\vel = 0,
\end{gathered}
\end{equation}
where $\vel$ is velocity, $\pres$ is pressure, $t$ is time, $\acc$ is acceleration due to external forces, $\visc$ is kinematic viscosity, and $\dens$ is density.

\paragraph{Vorticity-Based Monte Carlo}
\newcommand{\vort}{\omega}
\newcommand{\biotkernel}{B}
Let us recap the method by \mycitet{RiouxLavoie2022:Fluids} to highlight the challenges in developing a Monte Carlo solver for the velocity-based formulation in \cref{eq:ns}.
Their method uses the vorticity representation of the fluid. For instance, for inviscid fluid in 2D, the vorticity-based formulation simplifies the Navier-Stokes equations to the vorticity transport equation, $\frac{D\vort}{Dt}=0$, where $\frac{D}{Dt}$ is the material derivative.
We can write the velocity field used to advect the vorticity field as an integral, and combined with a semi-Lagrangian advection gives the equation they used to advance the vorticity field in each time step in the absence of obstacles:
\begin{equation*}%
     \vort(\vecx, t) \approx \vort \left(\vecx - \Delta t \int_{\mathbb{R}^2} \biotkernel(\vecx, \vecy) \times\vort(\vecy, t-\Delta t) \dVy, t-\Delta t\right),
\end{equation*}
where $\biotkernel(\vecx, \vecy)$ is the Biot-Savart kernel. They applied Monte Carlo integration to evaluate the integral.

The vorticity-based formulation offered a straightforward application of the Monte Carlo method. It is nontrivial to design a similar scheme for the velocity-based formulation in \cref{eq:ns} because the Navier-Stokes equations, in their velocity form, couple the velocity and the pressure variables in an intricate way: the pressure serves as a Lagrange multiplier that makes the velocity field incompressible. We address this challenge by adopting operator splitting. %

\paragraph{Operator splitting with pointwise estimators} The basic formulation of our method follows the standard operator splitting framework~\cite{Stam1999}. We solve \cref{eq:ns} by taking discrete time steps with step size $\Delta t$. For each time step, we decompose the Navier-Stokes equations into four substeps:
\begin{enumerate}
    \item advection: $\frac{\partial \vel}{\partial t} = - (\vel \cdot \grad)  \vel $,
    \item external force integration: $\frac{\partial \vel}{\partial t} = \acc$,
    \item diffusion: $\frac{\partial \vel}{\partial t} = \visc  \laplace \vel$, and
    \item projection: $\frac{\partial \vel}{\partial t} = - \frac{1}{\dens}\grad \pres$ such that $\diverg\vel=0$.
\end{enumerate}
In contrast to traditional discretization-based solvers, we develop a pointwise (Monte Carlo) estimator for each substep. We design a solver for each substep that can estimate the velocity at \emph{any} spatial point we are interested in, assuming we can likewise query the velocity estimates from the previous steps at any spatial point.

Basing our approach on this pointwise formulation offers unique advantages.
Our formulation is agnostic to the underlying discretization of the velocity field, which allows for flexible integration of velocity-based techniques used by traditional discretization-based solvers, including the common Stable Fluids-style~\cite{Stam1999} grid-based methods and grid-particle hybrid PIC/FLIP methods, into our Monte Carlo formulation.
While the vorticity-based Monte Carlo method~\mycite{RiouxLavoie2022:Fluids} has similar advantages of having flexible discretization options, their method cannot benefit from common velocity-based techniques.
Moreover, we will show in \cref{sec:results} that our formulation allows us to remove some of the excessive grid interpolation errors that are otherwise introduced in traditional solvers.%

Below, we formulate a pointwise estimate of the solution to each substep.
Later, in \cref{sec:results}, we will describe how we employ these estimators in a few different implementations, some using uniform grid velocity storage and others using hybrid grid-particle velocity storage.
To simplify notation, we use $\vel_0$ to $\vel_4$ to indicate the velocity field at each substep within a single time step. The initial velocity field for the current time step $\vel_0$ is defined as the output velocity field $\vel_4$ from the last time step. The advection step takes $\vel_0$ as its input and outputs $\vel_1$, and so on.

\subsection{Advection}\label{sec:advection}
We employ semi-Lagrangian advection, which estimates the new velocity at a point by tracing trajectories backward in time through the velocity field. For example, a forward Euler time discretization for the velocity at point $\vecx$ yields the update rule
\begin{equation}
    \vel_1(\vecx) \leftarrow \vel_0(\vecx - \dt\,\vel_0(\vecx)).
\end{equation}
This semi-Lagrangian update satisfies our pointwise evaluation requirement: we can estimate the velocity after the step by evaluating the input velocities at a few points. While the time-discretized equation above is identical to what appears in grid-based fluid solvers, we emphasize that the point where we evaluate the advected velocity, $\vecx$, does not necessarily need to be aligned with discretized locations (e.g., grid points), and when we evaluate the pre-advection velocity at points $\vecx$ and $\vecx - \dt\,\vel_0(\vecx)$, we are not restricted to evaluation by interpolation of some grid data. Thus, the pointwise form offers flexibility that is unavailable with classical solvers.
In practice, rather than forward Euler, we use a higher-order time integration scheme (third-order Runge-Kutta) to improve the accuracy of the traced trajectories, and the adaptations of error-correcting schemes such as MacCormack~\cite{Selle2008:MacCormack} and BFECC~\cite{Dupont2007:BFECC} are also trivial; all of these schemes likewise support pointwise evaluation. %

\subsection{External Force Integration}
In the presence of external forces leading to acceleration, such as a buoyancy force, we update the post-advection velocity using a forward Euler discretization:
\begin{equation}
     \vel_2(\vecx) \leftarrow \vel_1(\vecx) + \Delta t\, \acc(\vecx).
\end{equation}
This expression is again a pointwise update of the velocity field, where $\vecx$ is not necessarily at a discretized location.

\subsection{Projection}\label{sec:projection}
As the diffusion step is necessary only for viscous fluids and we can simply let $\vel_3\leftarrow\vel_2$ for inviscid fluids, we first discuss the projection step. The objective of the projection step is to find the pressure field that projects out the divergent velocity mode of the input field. The post-projection velocity at $\vecx$ is given by
\begin{equation}\label{eq:projvel}
 \vel_4(\vecx) \leftarrow \vel_3(\vecx) - \grad\ppres(\vecx),
\end{equation}
where $\ppres$ satisfies the Poisson equation
\begin{equation}\label{eq:pressure_poisson}
    \laplace\ppres(\vecx) = \diverg\vel_3(\vecx),
\end{equation}
and $\ppres = \frac{\dt}{\dens}\pres$. Since only $\grad\ppres$ is required for the update in \cref{eq:projvel}, the task is reduced to estimating the gradient of the solution of the Poisson equation; the value of $\ppres$ itself is not needed. 
In each subsection below, we discuss the integral formulation followed by its evaluation with Monte Carlo integration.

\subsubsection{Without solid boundaries}\label{sec:projection_wo_boundaries}
To simplify the problem, we will first consider a case without any solid boundaries (i.e., the domain is unbounded with no obstacles inside it).
Under this assumption, it is well-known that we can write the solution to the Poisson equation using the fundamental solution $\fund$ of the Laplace operator.  
The fundamental solution $\fund$ satisfies $\laplace \fund(\vecx, \vecy) + \delta(\vecx - \vecy) = 0$ in the unbounded domain, where $\delta$ is the Dirac delta function.
In 2D, $\fund(\vecx, \vecy) =  -\frac{1}{2\pi}\log r$, and in 3D, $\fund(\vecx, \vecy) = \frac{1}{4\pi r}$, where $r = \lVert \vecy - \vecx \rVert_2$.
We can then write the solution to \cref{eq:pressure_poisson} as a convolution of the source term of the Poisson equation and the fundamental solution:
\begin{equation}\label{eq:pressure_def}
\ppres(\vecx) = -\int_\Rd \fund(\vecx, \vecy)\, \grad_\vecy\cdot \vel_3(\vecy)\dVy,
\end{equation}
where dimension $\dimension=2, 3$. We used the subscript $\vecy$ to indicate that we perform differentiation with respect to $\vecy$, and we will use similar notation throughout this section.
By applying the Laplacian to \cref{eq:pressure_def}, one can confirm that $\ppres$ satisfies \cref{eq:pressure_poisson}.
To evaluate the gradient of $\ppres$, we take the gradient of \cref{eq:pressure_def}:
\begin{equation}\label{eq:pressure_grad}
    \grad_\vecx\ppres(\vecx) = -\int_\Rd \grad_\vecx\fund(\vecx, \vecy)\, \grad_\vecy\cdot \vel_3(\vecy)\dVy.
\end{equation}
This integral is still not suitable for numerical computation for our purposes, as we describe below. We therefore extend beyond the prior work by proposing a further transformation.

Since we typically consider storing information and performing volume integrals over a bounded simulation domain, we replace the (infinite) integral domain $\Rd$ in \cref{eq:pressure_grad} with a bounded simulation domain $\sdomain$ by assuming that the velocity divergence $\grad\cdot \vel_3$ is zero outside the simulation domain.

Attempting to use \cref{eq:pressure_grad} in this form for Monte Carlo integration still requires explicit evaluation of velocity divergence inside the domain, whereas we desire a solver that takes only a pointwise-evaluated velocity field as input.
There are a few possible approaches to obtain the required divergence. First, we could apply finite differences to the velocity field by accepting some errors. %
Second, we could differentiate the substep that precedes projection so that it outputs the necessary velocity divergence, in addition to the velocity itself. 
We propose instead a velocity-only design that we believe fits better in our Monte Carlo framework.

To eliminate the dependency on velocity divergence in \cref{eq:pressure_grad}, we use  the identity $\grad_\vecx\fund = -\grad_\vecy\fund$ and apply integration by parts:%
\begin{align}
\grad_\vecx\ppres(\vecx)\,
&\begin{aligned} = \int_\sdomain \left\{\grad_\vecy\fund(\vecx, \vecy)\right\} \grad_\vecy\cdot \vel_3(\vecy)\dVy\end{aligned}\\
&\begin{aligned} \label{eq:pressure_grad_vel_only}
       =&-\int_\sdomain \mathbf{H}(\vecx, \vecy)\,\vel_3(\vecy) \dVy\\
        &- \int_\sboundary \left\{\grad_\vecx\fund(\vecx, \vecy)\right\}\vecn(\vecy)\cdot\vel_3(\vecy) \dAy,\\  
\end{aligned}
\end{align}
where $\vecn$ is the outward unit normal. 
The Hessian of the fundamental solution denoted $\mathbf{H}$, is given by
\begin{align}\label{eq:hessian}
\mathbf{H}(\vecx, \vecy) &= \mathbf{S}(\vecx, \vecy) - \frac{\delta(\vecr)}{\dimension}\identity, &
\mathbf{S}(\vecx, \vecy) &=  \frac{1}{\lvert\partial B\rvert r^{\dimension+2}}(\dimension\vecr\vecr\transpose - r^2\identity),
\end{align}
where $\lvert\partial B\rvert$ is the surface area of a unit $(\dimension-1)$-sphere, $\vecr = \vecy - \vecx$, ${r=\lVert\vecr\rVert_2}$ and $\identity$ is the identity matrix. While we use the notation above for ease of understanding, technically, it needs to be understood in the sense of the generalized (distributional) derivative, and interested readers can refer to the discussion by \citet{Frahm1983} and \citet{Hnizdo2011} for further details. Substituting \cref{eq:hessian} into \cref{eq:pressure_grad_vel_only}, we get%
\begin{equation}\label{eq:pressure_grad_final}
\begin{aligned}
\grad_\vecx\ppres(\vecx) =& -\int_\sdomain \mathbf{S}(\vecx, \vecy) \,\vel_3(\vecy) \dVy + \frac{1}{\dimension}\vel_3(\vecx)   \\
&- \int_\sboundary \left\{\grad_\vecx\fund(\vecx, \vecy)\right\}\vecn(\vecy)\cdot\vel_3(\vecy)\dAy\\
 =&  -\int_\sdomain \mathbf{S}(\vecx, \vecy) \left\{\vel_3(\vecy) - \vel_3(\vecx)\right\} \dVy \\
&- \int_\sboundary \left\{\grad_\vecx\fund(\vecx, \vecy)\right\}\vecn(\vecy)\cdot\left\{\vel_3(\vecy)-\vel_3(\vecx)\right\} \dAy.
\end{aligned}
\end{equation}
Again, technically, the domain of the first integral should exclude an infinitesimal ball around $\vecx$ \cite{Hnizdo2011}, but we use this simple notation for readability.
In \cref{eq:pressure_grad_final}, to get the final expression, we replaced the original velocity field with a velocity field that is globally shifted by the constant velocity at point $\vecx$, $\vel_3(\vecx)$. The computed pressure gradient remains unchanged because the divergence of the shifted velocity field is the same as the original one. This global shift cancels the zeroth order term of the Taylor expansion of \(\mathbf{u}_3(\mathbf{y})\) around \(\mathbf{x}\), and the remaining terms of the Taylor expansion multiplied by the function $\mathbf{S}$ will have 
a singularity of $O(1/r^{\dimension-1})$ instead of the original $O(1/r^\dimension)$ as $r\rightarrow 0$. This lower-order singularity can be handled by an appropriate importance-sampling strategy, as we discuss next.

\paragraph{Monte Carlo Estimation}
Now that we have an integral representation of the pressure gradient, we can define a Monte Carlo estimator for the pressure gradient. We sample $N_V$ points $\vecy^i$ inside the simulation domain $\sdomain$ and $N_A$ points $\vecy^j$ on the simulation domain boundary $\sboundary$ using sampling strategies with probability density functions (PDFs) $\pdf_V$ and $\pdf_A$, respectively, to get
\begin{equation}\label{eq:mc_pressure_grad}
\grad_\vecx\ppres(\vecx) \approx  \estimator{E_V(\vecx)} + \estimator{E_A(\vecx)}
\end{equation}
where
\begin{align}
    \label{eq:EV}\estimator{E_V(\vecx)}&= -\frac{1}{N_V}\sum_{i=1}^{N_V} \frac{\mathbf{S}(\vecx, \vecy^i)}{\pdf_V(\vecy^i|\vecx)} \left\{\vel_3(\vecy^i) - \vel_3(\vecx)\right\} \\
    \label{eq:EA}\estimator{E_A(\vecx)}&= - \frac{1}{N_A}\sum_{j=1}^{N_A}  \frac{\grad_\vecx\fund(\vecx, \vecy^j)}{\pdf_A(\vecy^j|\vecx)}\,\vecn(\vecy^j)\cdot\left\{\vel_3(\vecy^j)-\vel_3(\vecx)\right\}.
\end{align}
This formulation is an unbiased estimator for the pressure gradient if the probability that we sample any point with a nonzero contribution is nonzero. If the integrand is non-negative everywhere, importance sampling according to the PDF that is roughly proportional to the integrand can decrease the variance of the estimator. While our integrands may contain negative values, we follow this idea to design our sampling strategy. 
In our implementation, for \cref{eq:EV}, to handle the singular integral, we draw samples such that PDF $\pdf_V$ is proportional to $1/r^{(\dimension-1)}$ inside of the smallest ball around $\vecx$ that fully contains the entire simulation domain, and set zero contributions from the samples outside the simulation domain. This approach is equivalent to evaluating the integral we get by extending the original integral domain to the ball and putting zero integrands in the extended part. We also use antithetic sampling here: in addition to a point $\vecy^i$ sampled this way, we always sample the symmetric point with respect to $\vecx$, $2\vecx - \vecy^i$ to further reduce variance for smooth velocity fields.
For \cref{eq:EA}, we uniformly sample points on the simulation boundary. Properly sampling the positive and negative contributions separately may further reduce the variance of the estimators~\cite[Section 9.12]{Owen:MonteCarlo, Chang2023:ReSTIRDR}, but we leave it as future work. 
Equation \ref{eq:mc_pressure_grad} only requires the ability to evaluate the velocity at the position of sample points to perform projection, in contrast to traditional solvers that typically require a globally coupled linear system solve.

\subsubsection{With solid boundaries.} \label{sec:projection_with_boundaries}
\begin{figure}[b]
    \centering
    \includegraphics[angle=90, origin=c, trim={1cm 0 0 0}, clip, width=0.18\linewidth]{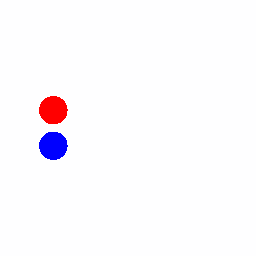}
    \includegraphics[angle=90, origin=c, trim={1cm 0 0 0}, clip, width=0.18\linewidth]{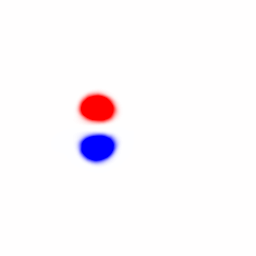}
    \includegraphics[angle=90, origin=c, trim={1cm 0 0 0}, clip, width=0.18\linewidth]{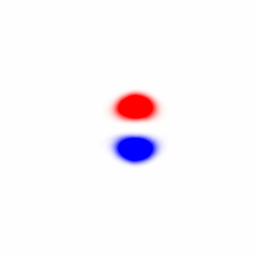}
    \includegraphics[angle=90, origin=c, trim={1cm 0 0 0}, clip, width=0.18\linewidth]{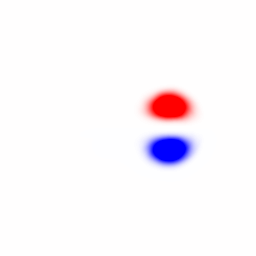}
    \includegraphics[angle=90, origin=c, trim={1cm 0 0 0}, clip, width=0.18\linewidth]{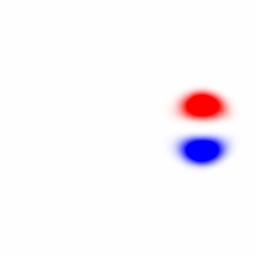}
\vspace{.5pt}
    \hrule
\vspace{1pt}
    \includegraphics[angle=90, origin=c, trim={1cm 0 0 0}, clip, width=0.18\linewidth]{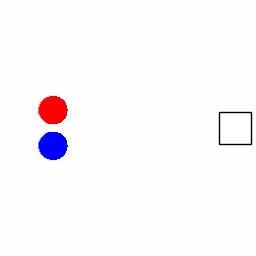}
    \includegraphics[angle=90, origin=c, trim={1cm 0 0 0}, clip, width=0.18\linewidth]{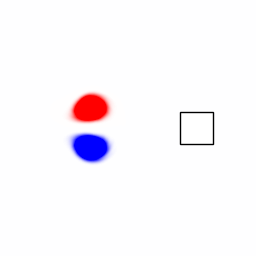}
    \includegraphics[angle=90, origin=c, trim={1cm 0 0 0}, clip, width=0.18\linewidth]{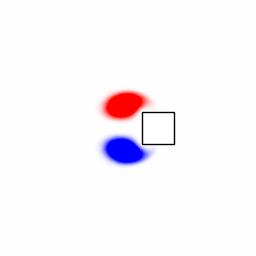}
    \includegraphics[angle=90, origin=c, trim={1cm 0 0 0}, clip, width=0.18\linewidth]{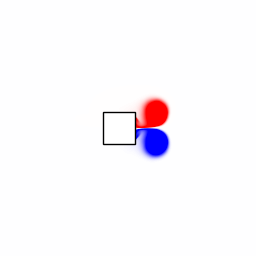}
    \includegraphics[angle=90, origin=c, trim={1cm 0 0 0}, clip, width=0.18\linewidth]{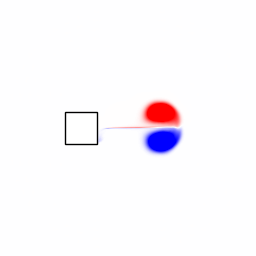}
    \hrule
\vspace{1pt}
    \includegraphics[angle=90, origin=c, width=0.18\linewidth]{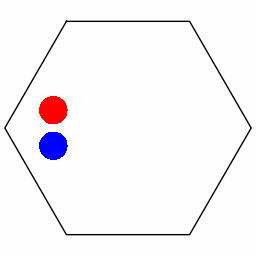}
    \includegraphics[angle=90, origin=c, width=0.18\linewidth]{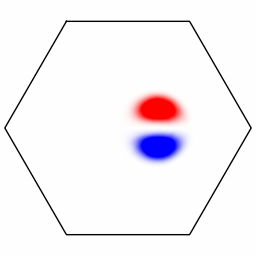}
    \includegraphics[angle=90, origin=c, width=0.18\linewidth]{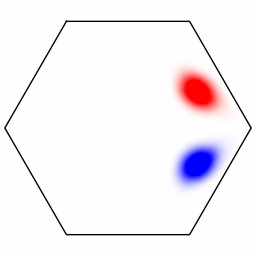}
    \includegraphics[angle=90, origin=c, width=0.18\linewidth]{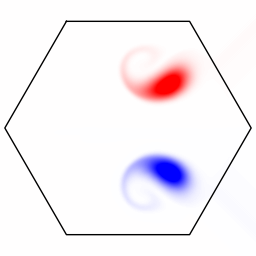}
    \includegraphics[angle=90, origin=c, width=0.18\linewidth]{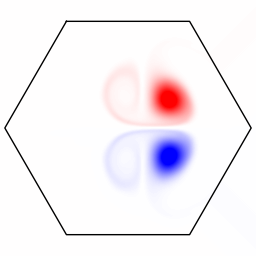}
    
    \caption{Our method can correctly simulate scenes with different boundary types (resolution $256^3$). From the left, we simulate the colored smoke advected with the velocity field induced by vortex pairs moving with no obstacles (left), in an unbounded domain with a moving square obstacle (middle), and in a domain bounded by an obstacle (right). For each simulation, we show the time evolution from the left to the right, but the time stamps differ for each setup.\hfill\videotime{1}{17}}
    \label{fig:boundary}
\end{figure}

Next, we consider the case when solid object boundaries are involved (\cref{fig:boundary}).
We still solve \cref{eq:pressure_poisson} for $\grad\ppres$, but with free slip boundary conditions, 
\begin{equation}\label{eq:boundary_cond}
    \frac{\partial \ppres}{\partial \vecn} = \vecn\cdot (\vel_3 - \velsol)
\end{equation}
on solid boundaries, where $\vecn$ is the unit outward normal from the fluid domain $\Omega$, $\frac{\partial}{\partial \vecn}=\vecn\cdot\grad$ is a normal derivative, and $\velsol$ is the solid velocity. If the simulation domain is bounded, this problem is an interior Neumann problem for the Poisson equation, and otherwise, it is an exterior Neumann problem; we need a Monte Carlo solver that is capable of handling these problems.

The basic walk-on-spheres method~\cite{Sawhney:2020:MCG} can only handle Dirichlet problems, and its applicability to unbounded domain problems leaves some concerns about the termination of paths without additional bias.  \mycitet{RiouxLavoie2022:Fluids} arbitrarily terminated their walk-on-spheres paths after a few steps in their simulation, leaving some additional bias. \citet{Nabizadeh2021} suggested inverting the unbounded domain into a bounded one using the Kelvin transform so that paths always terminate, but using this approach would require very different treatment for bounded and unbounded domains.

To solve Neumann problems, the walk-on-stars method \mycite{sawhney2023walk} extended the walk-on-spheres method with carefully-designed recursion relationships and tailored sampling techniques based on the extensions by \citet{Simonov2008} and \citet{Ermakov2009}.
For a Neumann problem in an unbounded domain, however, the Kelvin transform \cite{Nabizadeh2021} is still required; it transforms a Neumann problem into a Robin problem, to which the application of walk-on-stars has not yet been attempted.
Among others, the walk-on-boundary method~\cite{sabelfeld1982vector, SabelfeldSimonov1994}, introduced recently to graphics by \citet{Sugimoto2023}, can be applied to Neumann boundary problems in both bounded and unbounded domains under a unified framework. The method builds upon a boundary integral equation analogous to the rendering equation for light transport simulation~\cite{Kajiya:1986:Rendering}, and uses ray-tracing to solve the boundary integral equation, and we use this method in our solver. %

Following \citet{SabelfeldSimonov1994}, we can write the gradient of the solution to the Poisson equation of \cref{eq:pressure_poisson} with the boundary condition of \cref{eq:boundary_cond} using the single layer potential formulation and define an integral equation for the unknown density function.
For our application, these equations contain the volume integral terms arising from the source term in the Poisson equation of \cref{eq:pressure_poisson} specific to our problem. 
Similar to \cref{sec:projection_wo_boundaries}, we transform these volume integral terms as follows.
If the domain $\domain$ is unbounded, we suggest replacing it with a bounded simulation domain $\sdomain$, assuming zero velocity divergence outside of the simulation domain; otherwise, we let $\sdomain=\domain$. With this definition, we have  $\boundary \subseteq \sboundary$ for the boundaries, and we distinguish these two sets of boundaries in the following equations. We also remove the explicit dependencies on the velocity divergence by integration by parts similarly to \cref{eq:pressure_grad_vel_only} to get%
\begin{equation}\label{eq:wob_pressure_grad_vel_only}
\begin{aligned}
    \grad_\vecx\ppres(\vecx) =
    & \int_\boundary \left\{\grad_\vecx\fund(\vecx, \vecy)\right\} \left[\wobdens(\vecy) - \vecn(\vecy)\cdot \left\{\vel_3(\vecy) - \vel_3(\vecx)\right\}\right]\dAy \\
    &-\int_\sdomain \mathbf{S}(\vecx, \vecy) \left\{\vel_3(\vecy) - \vel_3(\vecx)\right\} \dVy\\
    &-\int_{\sboundary\backslash\boundary} \left\{\grad_\vecx\fund(\vecx, \vecy)\right\}  \vecn(\vecy)\cdot \left\{\vel_3(\vecy) - \vel_3(\vecx)\right\}\dAy \\
\end{aligned}
\end{equation}
for $\vecx \in \domain$ and 
\begin{equation}\label{eq:wob_density_grad_vel_only}
\begin{aligned}
    \wobdens(\vecx) =
    &  -\int_\boundary 2\dGdnx(\vecx, \vecy) \left[\wobdens(\vecy) - \vecn(\vecy)\cdot \left\{\vel_3(\vecy) - \vel_3(\vecx)\right\}\right]\dAy \\
    &+\int_\sdomain 2\vecn(\vecx)\transpose\mathbf{S}(\vecx, \vecy) \left\{\vel_3(\vecy) - \vel_3(\vecx)\right\} \dVy\\
    &+\int_{\sboundary\backslash\boundary} 2\dGdnx(\vecx, \vecy)\,  \vecn(\vecy)\cdot \left\{\vel_3(\vecy) - \vel_3(\vecx)\right\}\dAy \\
    & + 2\vecn(\vecx)\cdot \{\vel_3(\vecx) - \velsol(\vecx)\}
\end{aligned}
\end{equation}
for $\vecx \in \boundary$.
In both \cref{eq:wob_pressure_grad_vel_only} and \cref{eq:wob_density_grad_vel_only}, we have the boundary integral terms that involve the unknown density function $\wobdens$ on the solid boundaries $\boundary$, and all the other integrals involve only known quantities.
Note \cref{eq:wob_pressure_grad_vel_only} is a generalization of the strategy described in \cref{sec:projection_wo_boundaries} as we can recover \cref{eq:pressure_grad_final} by dropping the solid boundary integral term.%

\paragraph{Monte Carlo Estimation}
Based on \cref{eq:wob_pressure_grad_vel_only} and \cref{eq:wob_density_grad_vel_only}, we get a biased walk-on-boundary Monte Carlo estimator~\cite{SabelfeldSimonov1994, Sugimoto2023}.
Compared to the simplest formulation for the Laplace equation~\cite{Sugimoto2023}, we have some additional non-recursive terms in \cref{eq:wob_pressure_grad_vel_only} and \cref{eq:wob_density_grad_vel_only} as a result of the transformations discussed above; we must consider how to sample these added terms. We choose to sample the non-recursive contributions on $\partial \domain$ using ray intersection sampling for importance sampling and estimate the other non-recursive terms similarly to \cref{sec:projection_wo_boundaries}. To improve efficiency, we additionally employ boundary value caching, which lets us share the walk-on-boundary subpaths among evaluation points akin to the virtual point light (VPL) method~\cite{keller1997instant} in rendering. We describe the details of our particular sampling strategy in the supplemental note. Our Monte Carlo method is compatible with a variety of sampling techniques, so further variance reduction is likely possible. %

\subsection{Diffusion}
Simulating viscous fluids requires an additional diffusion step before the projection step. The set of equations we solve here is the constant-coefficient diffusion equation, %
\begin{equation}\label{eq:originaldiffusioneq}
    \begin{aligned}
        \frac{\partial \overline\vel(\vecx, s)}{\partial s} &= \visc  \laplace \overline\vel(\vecx, s)  & \text{for}&\; \vecx \in \domain, s \in (0, \Delta t),\\
        \overline\vel(\vecx, s) &= \vel_s(\vecx) & \text{for}&\; \vecx \in \boundary, s \in (0, \Delta t),\quad\text{and}\\
        \overline\vel(\vecx, 0) &= \vel_2(\vecx) & \text{for}&\; \vecx \in \domain,\\
    \end{aligned}
\end{equation}
and the output of this diffusion step is $\vel_3(\vecx) = \overline\vel(\vecx, \Delta t)$. 
The time $s$ here represents the time within each diffusion time step. We use the overbar to indicate that the velocity variable $\overline\vel$ takes time $s$ as defined here, and differs from the other sections.
The solid velocity gives the no-slip boundary condition, and the output from the preceding advection step gives the initial conditions.

When there are no solid boundaries or the time step size is small enough to ignore boundary effects, we can omit the second equation describing the boundary condition. In such a case, evaluating a convolution of the input velocity field with a Gaussian function would give the exact solution to the diffusion equation. 
Though they apply it to the vorticity field rather than the velocity field, \mycitet{RiouxLavoie2022:Fluids} use such a convolution-based approach to model diffusion.
This approach fails when either the time step size is relatively large or the boundary effect is important.
Traditional grid-based solvers similarly cannot rely on a simple Gaussian filter in the presence of large time steps or obstacles, and must instead solve a globally coupled linear system for viscosity \cite{Bridson2015}.\looseness=-1

To address this problem, we utilize the walk-on-boundary method for the diffusion equation~\cite[Chapter 4]{SabelfeldSimonov1994} to solve \cref{eq:originaldiffusioneq}. Similarly to \cref{sec:projection_with_boundaries}, the diffusion walk-on-boundary method is a pointwise estimator and is a natural generalization of the case when there are no boundaries. One difference is that the boundary integral equation and the walk-on-boundary steps are now defined in the space-time domain, and we must sample paths from the point where we want to evaluate the solution with time $\Delta t$ towards the initial time within the time step, in the negative time direction.

Notably, for the diffusion walk-on-boundary, the recursion depth does not need to be predefined. By sampling the space-time paths using a PDF proportional to the integral kernel of the integral equation for the diffusion equation, we can terminate recursions when the sampled time is negative.
In our experiments, we observed that the diffusion step is cheaper than the projection step, and we did not attempt any algorithmic improvements, such as the reuse of subpaths. However, there remain many choices for sampling and various efficiency improvement strategies should apply here, too.
The supplemental material summarizes the diffusion walk-on-boundary method for readers' convenience.

\section{Results}\label{sec:results}
\begin{figure*}
\centering
\includegraphics[width=0.16\linewidth]{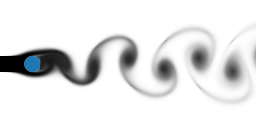}  
\includegraphics[width=0.16\linewidth]{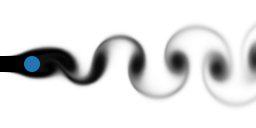}  
\includegraphics[width=0.16\linewidth]{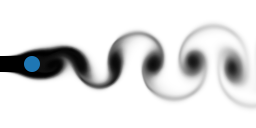} 
\includegraphics[width=0.16\linewidth]{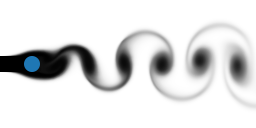}
\includegraphics[width=0.16\linewidth]{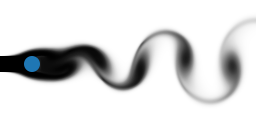}
\includegraphics[width=0.16\linewidth]{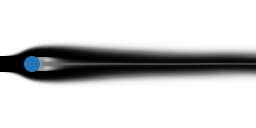}

\begin{minipage}[t]{0.16\linewidth}\centering\textsf{\small(a) Traditional grid-based\looseness-1\\ \cite{Batty2007}}\end{minipage}
\begin{minipage}[t]{0.16\linewidth}\centering\textsf{\small(b) Caching after projection and advection}\end{minipage}\centering
\begin{minipage}[t]{0.16\linewidth}\centering\textsf{\small(c) No caching after\\projection}\end{minipage}
\begin{minipage}[t]{0.16\linewidth}\centering\textsf{\small(d) No caching after\\advection}\end{minipage}
\begin{minipage}[t]{0.16\linewidth}\centering\textsf{\small(e) $\textit{Re} = 250\, (\nu=0.001)$}\end{minipage}
\begin{minipage}[t]{0.16\linewidth}\centering\textsf{\small(f) $\textit{Re} = 25\, (\nu=0.01)$}\end{minipage}
       
\begin{minipage}[t]{0.16\linewidth}\;\end{minipage}
\includegraphics[width=0.16\linewidth]{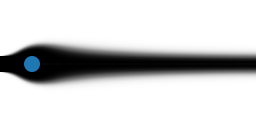}
\includegraphics[width=0.16\linewidth]{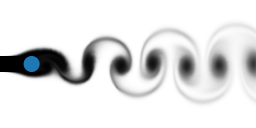}
\includegraphics[width=0.16\linewidth]{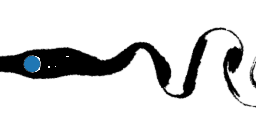}
\includegraphics[width=0.16\linewidth]{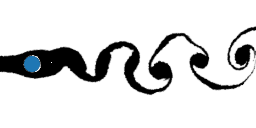}
\includegraphics[width=0.16\linewidth]{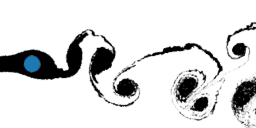}
\begin{minipage}[t]{0.16\linewidth}\;\end{minipage}
\begin{minipage}[t]{0.16\linewidth}\centering\textsf{\small(g) $\textit{Re} = 2.5\, (\nu=0.1)$}\end{minipage}
\begin{minipage}[t]{0.16\linewidth}\centering\textsf{\small(h) Advection-Reflection}\end{minipage}
\begin{minipage}[t]{0.16\linewidth}\centering\textsf{\small(i) PIC}\end{minipage}
\begin{minipage}[t]{0.16\linewidth}\centering\textsf{\small(j) FLIP 0.95}\end{minipage}
\begin{minipage}[t]{0.16\linewidth}\centering\textsf{\small(k) FLIP}\end{minipage}
        
    \caption{Variants of our method with different configurations tested on the K\'{a}rm\'{a}n vortex street scene with resolution $128\times256$. Except for (a), all results are generated with our Monte Carlo method. (b) to (d) compare the caching alternatives, (e) to (g) demonstrate simulations with various Reynolds numbers $\textit{Re}$, (h) shows the application of the advection-reflection method to yield reduced numerical dissipation, and (i) to (k) show the application of PIC/FLIP methods. Note that (i) to (k) have no density diffusion because they employ density advection by particles. Given their agreement with the result of the traditional grid-based method (a), we consider all variants of our method to produce qualitatively reasonable results.\hfill\videotime{1}{56}}
    \label{fig:vortexstreet}
\end{figure*}

We explain the results of our practical implementation of the method with discretization structures and other possible extensions.
We implemented our method using CUDA for GPU parallelism and the NVIDIA OptiX ray tracing engine~\cite{Parker2010} via the OWL wrapper library~\cite{owl2023} to accelerate the ray intersection queries.
For all 2D results, we use consistent parameters as listed in the supplemental note except for \cref{fig:2dbunny}.
The figures have time stamps for the corresponding animations in the supplemental video.\looseness=-1

\paragraph{Caching of velocity fields}
While our fundamental formulation is agnostic to a choice of intermediate storage of velocity fields, or absence of it by the expensive recursive application of estimators to the initial time step (see \cref{sec:discussion}), our basic implementation uses a simple uniform grid structure to store the velocities at grid nodes at each time step, and we refer to such a strategy a caching strategy in contrast to the fully time-recursive (cacheless) alternative. %
Since we do not evaluate any finite differences on the stored data, we do not need to use a staggered grid~\cite{Harlow1965}.
When we query a velocity from the cache, we bilinearly or trilinearly interpolate the values, and for points outside the cached domain, we take the velocity at the nearest cache point. As with classical grid-based solvers, interpolation- and advection-induced errors sometimes cause us to query values from points inside of solid obstacles; for this issue, we assign velocities inside solid obstacles to take velocities from the solid itself, giving no-slip behavior on the boundary.
Alternatively, one could fill the velocities inside solid obstacles with some form of extrapolation to approximate free slip behavior, for example, but we use the first approach throughout this paper.%

While we could cache the velocity field after each substep, the pointwise estimation capability of our approach lets us avoid some of the excessive velocity caching between substeps and reduce the associated errors. For example, we need no caching between the advection and the following projection step for inviscid simulation if we query the advected velocity values directly on the fly whenever the projection estimator needs such a pointwise input velocity. Alternatively, we can also design a method without velocity caching between the projection and the next advection so the velocity field is always advected according to the pointwise divergence-free velocity field.
We implemented these two options in addition to the option where we cache the resulting velocity field both after advection and after projection in \cref{fig:vortexstreet} (b) to (d). Our experiments showed consistent results among the three options; further investigations are needed to understand when having fewer caching errors can make critical differences.%

\paragraph{Boundary conditions} \cref{fig:boundary} shows that our method can correctly simulate the velocity field due to vortex pairs under different boundary conditions. In \cref{fig:teaser}, we further show our method's application to a domain with two disjoint obstacles. In such cases, vorticity-based methods typically use an incorrect boundary condition, and the vorticity-based Monte Carlo method~\mycite{RiouxLavoie2022:Fluids} fails to simulate the physics correctly. Applying the modification proposed by \citet{Yin2023} for vorticity-based simulation to a Monte Carlo method is not straightforward and has not been demonstrated. Hence, ours is currently the only Monte Carlo method that can produce correct simulation results in these scenarios. While the relative performance depends heavily on parameter choices, for the specific simulations in \cref{fig:boundary}, ours took 101.0s per time step while the vorticity-based method took 87.6s, with a difference of about 15\%.

\begin{figure}[b]
    \centering
    \includegraphics[trim={2cm 1cm 2cm 0.5cm}, clip, height=0.28\linewidth]{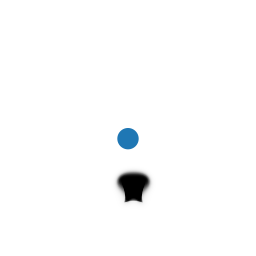}
    \includegraphics[trim={2cm 1cm 2cm 0.5cm}, clip, height=0.28\linewidth]{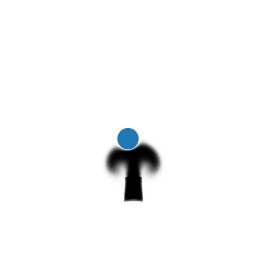}
    \includegraphics[trim={1.5cm 1cm 1.5cm 0.5cm}, clip, height=0.28\linewidth]{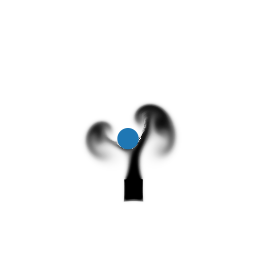}
    \includegraphics[trim={1.5cm 1cm 1.5cm 0.5cm}, clip, height=0.28\linewidth]{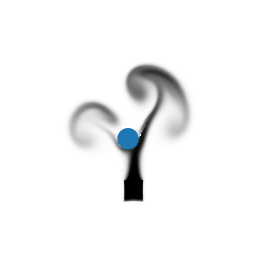}
    \includegraphics[trim={1cm 1cm 1cm 0.5cm}, clip, height=0.28\linewidth]{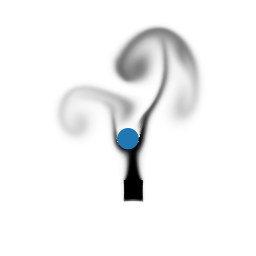}\\
    \includegraphics[trim={2cm 1cm 2cm 0.5cm}, clip, height=0.28\linewidth]{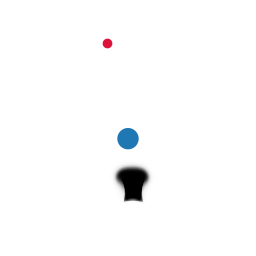}
    \includegraphics[trim={2cm 1cm 2cm 0.5cm}, clip, height=0.28\linewidth]{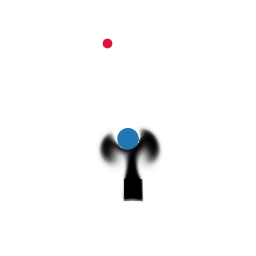}
    \includegraphics[trim={1.5cm 1cm 1.5cm 0.5cm}, clip, height=0.28\linewidth]{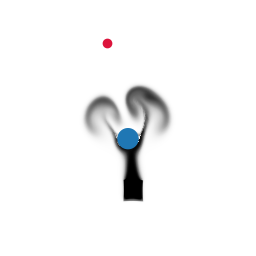}
    \includegraphics[trim={1.5cm 1cm 1.5cm 0.5cm}, clip, height=0.28\linewidth]{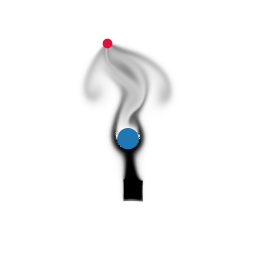}
    \includegraphics[trim={1cm 1cm 1cm 0.5cm}, clip, height=0.28\linewidth]{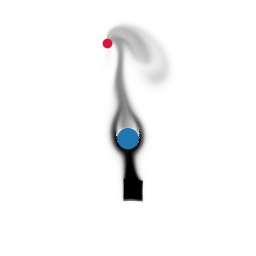}\\
    \caption{Buoyancy and divergence control (resolution $256^2$). We simulate a hot smoke plume rising towards a blue circular obstacle in 2D using the Boussinesq buoyancy model (left). We can also add a velocity sink to the scene, indicated with a red dot, causing the smoke to be sucked into the sink (right).\hfill\videotime{3}{49}}
    \label{fig:buoyancy}
\end{figure}

\begin{figure*}
    \includegraphics[width=0.16\linewidth]{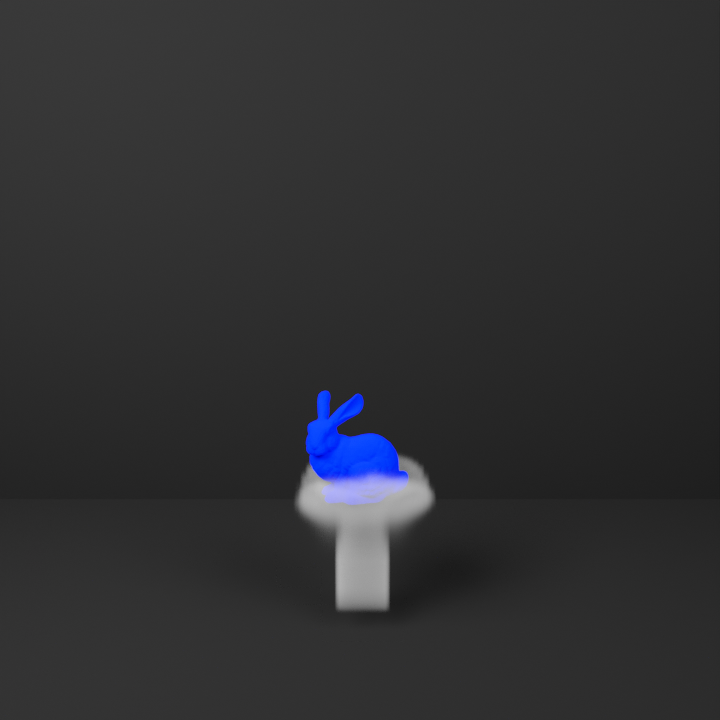}
    \includegraphics[width=0.16\linewidth]{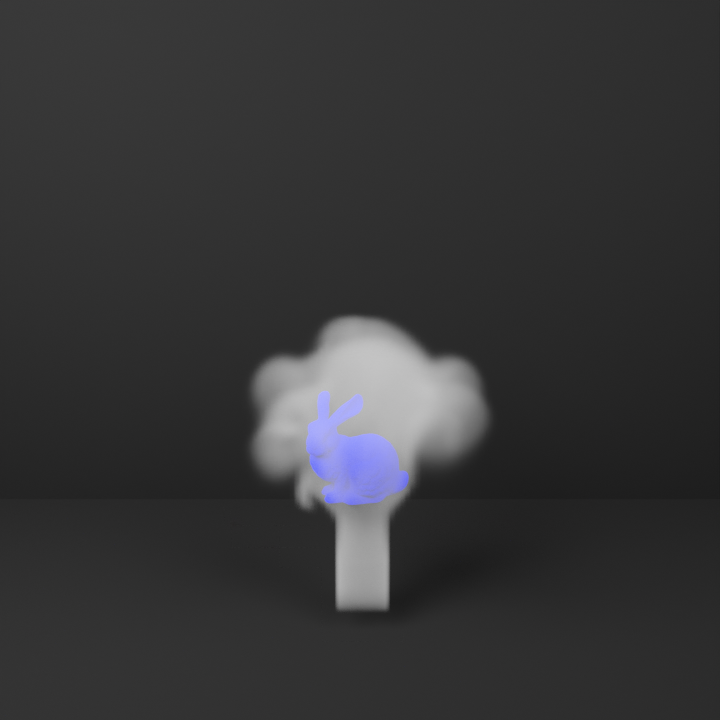}
    \includegraphics[width=0.16\linewidth]{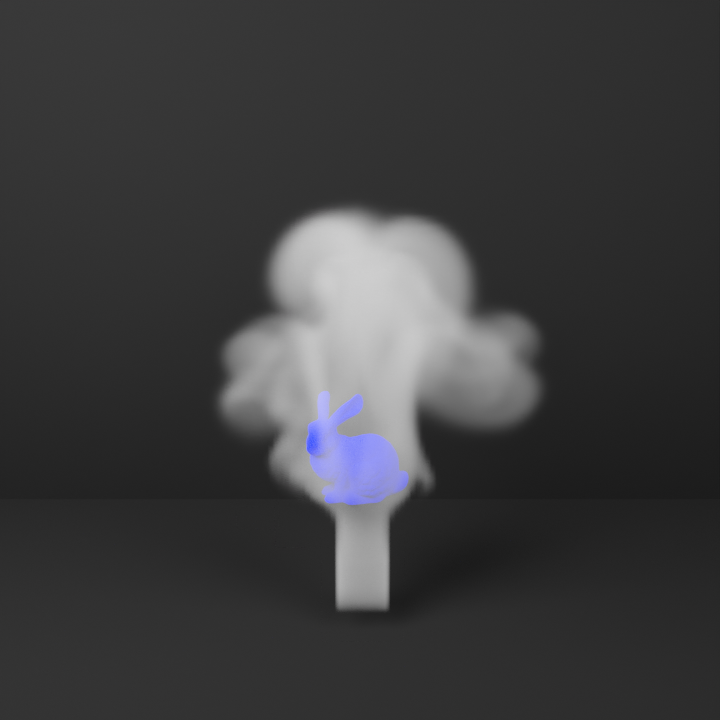}
    \includegraphics[width=0.16\linewidth]{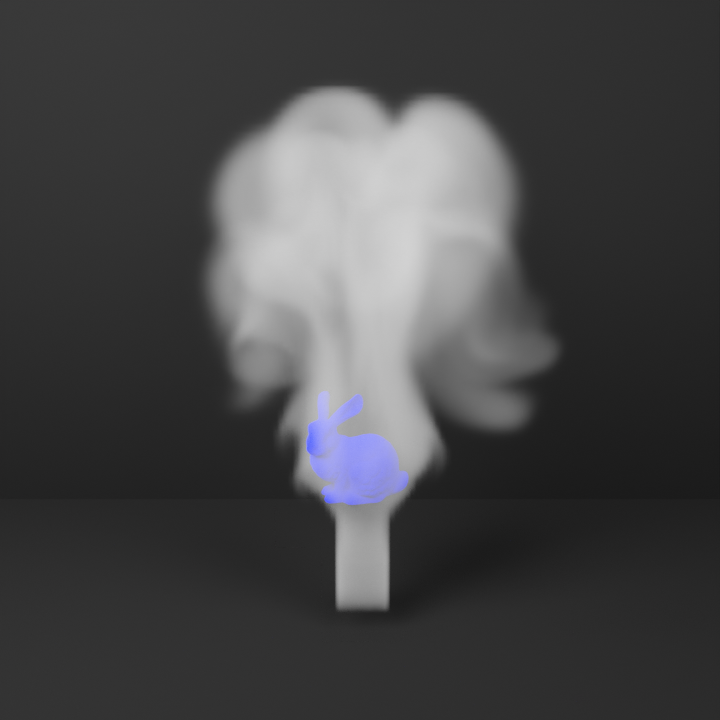}
    \includegraphics[width=0.16\linewidth]{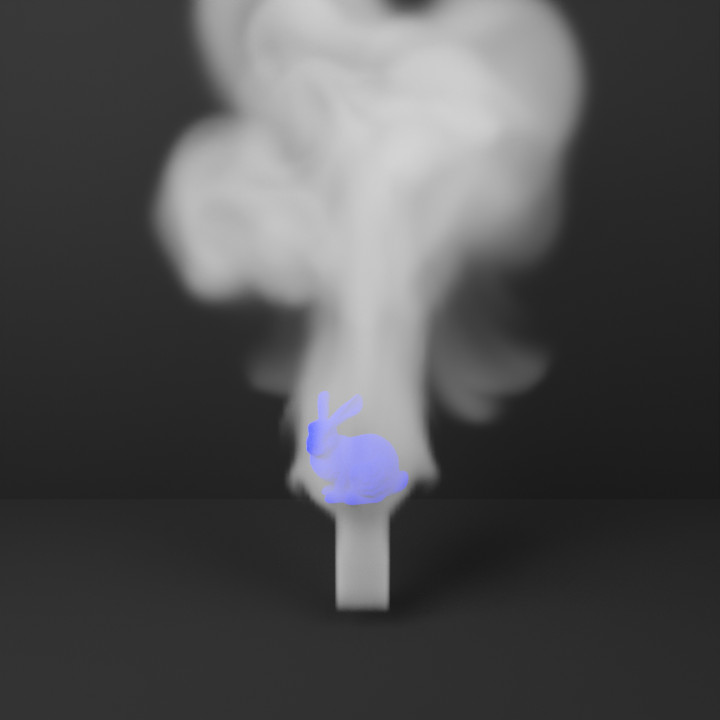}
    \includegraphics[width=0.16\linewidth]{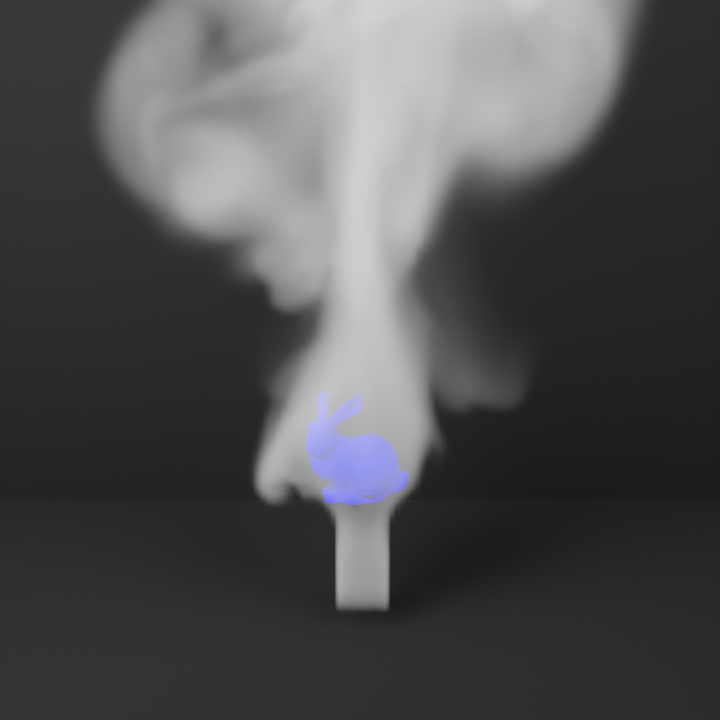}\\
    
    \includegraphics[width=0.16\linewidth]{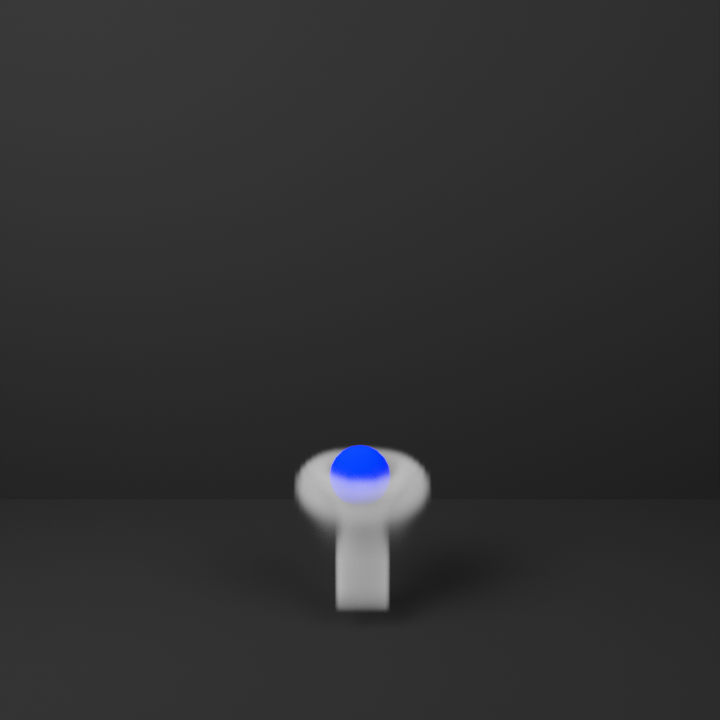}
    \includegraphics[width=0.16\linewidth]{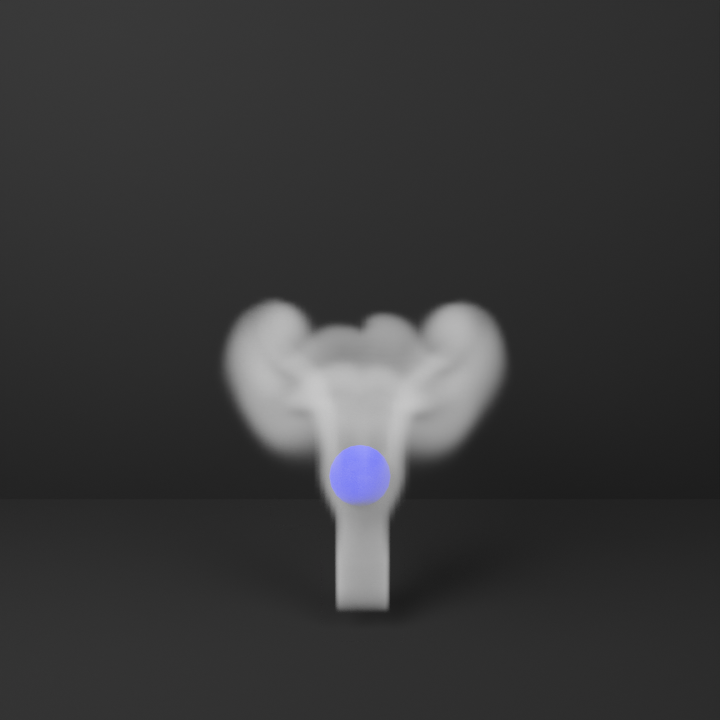}
    \includegraphics[width=0.16\linewidth]{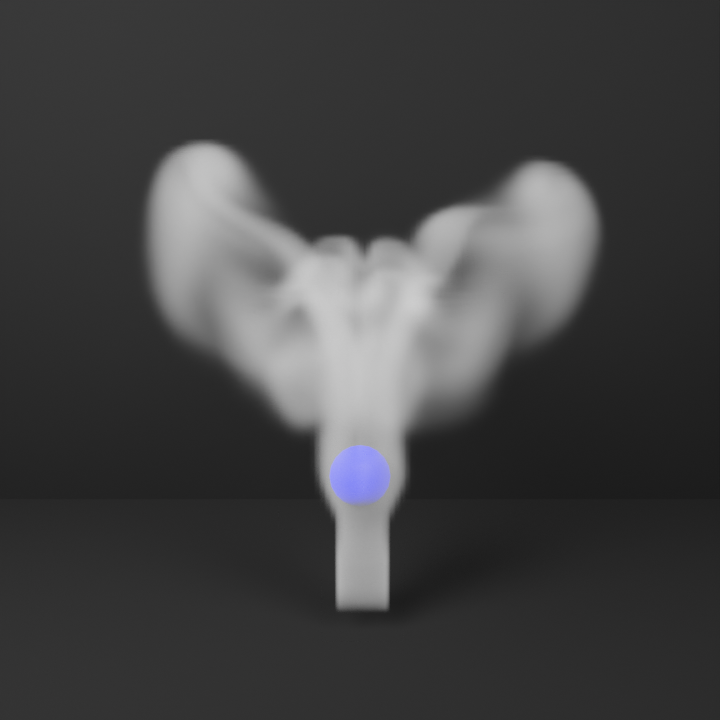}
    \includegraphics[width=0.16\linewidth]{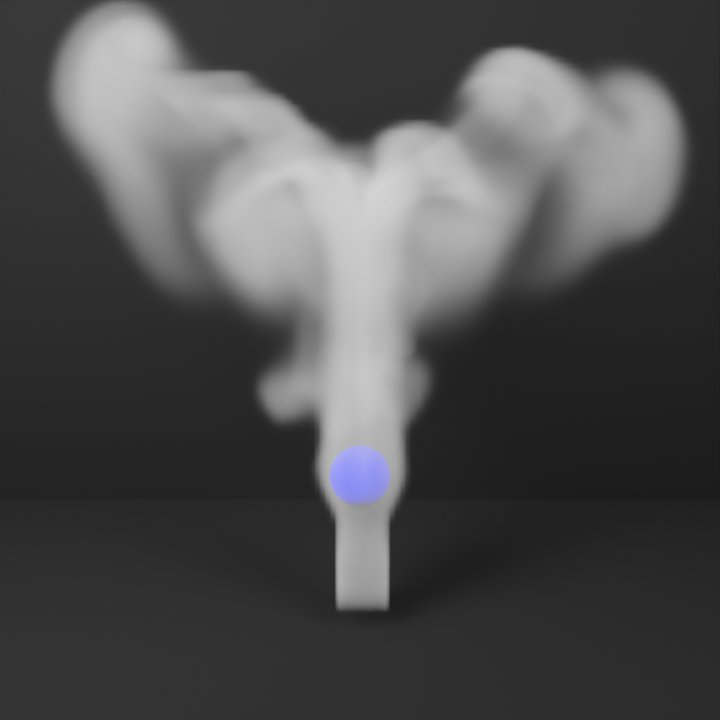}
    \includegraphics[width=0.16\linewidth]{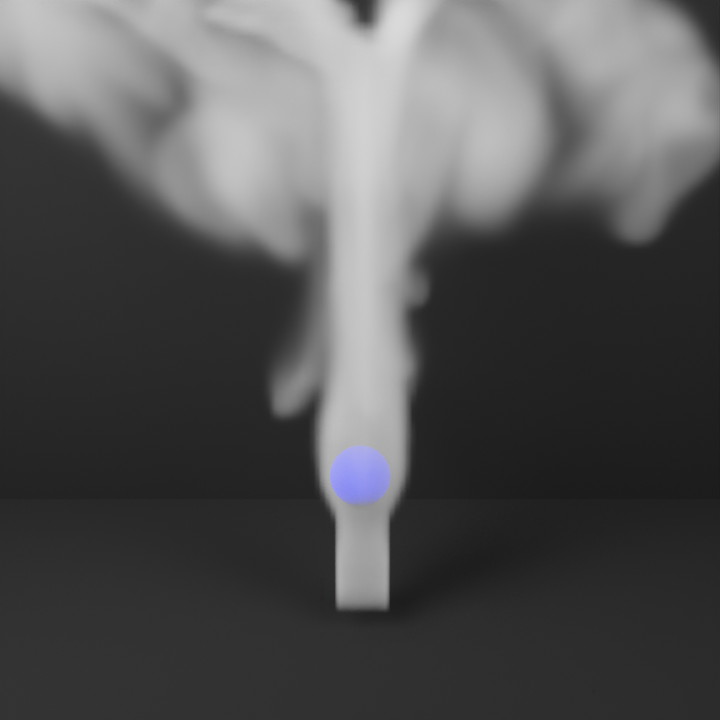}
    \includegraphics[width=0.16\linewidth]{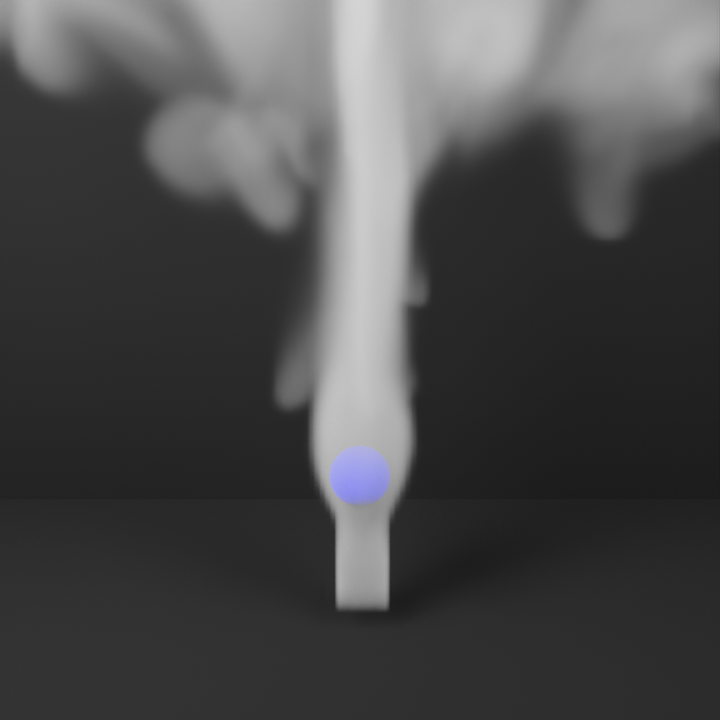}\\

    \includegraphics[width=0.16\linewidth]{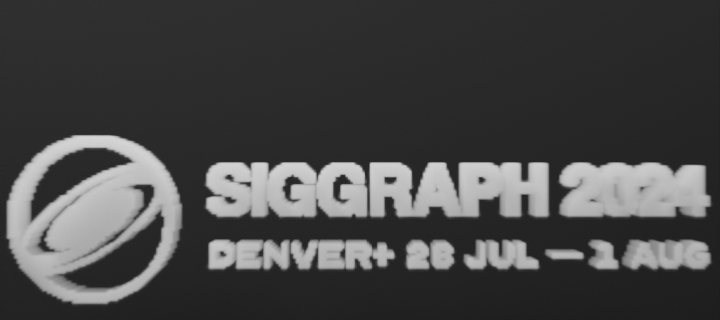}
    \includegraphics[width=0.16\linewidth]{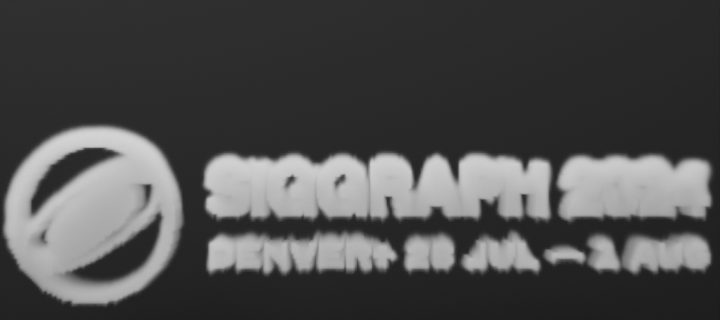}
    \includegraphics[width=0.16\linewidth]{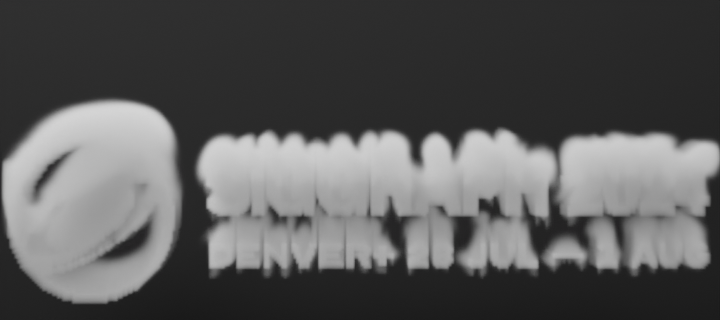}
    \includegraphics[width=0.16\linewidth]{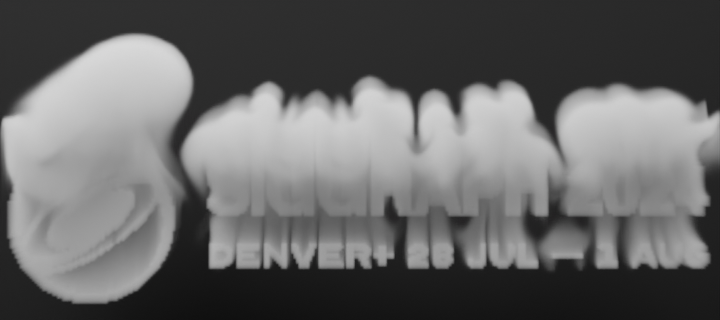}
    \includegraphics[width=0.16\linewidth]{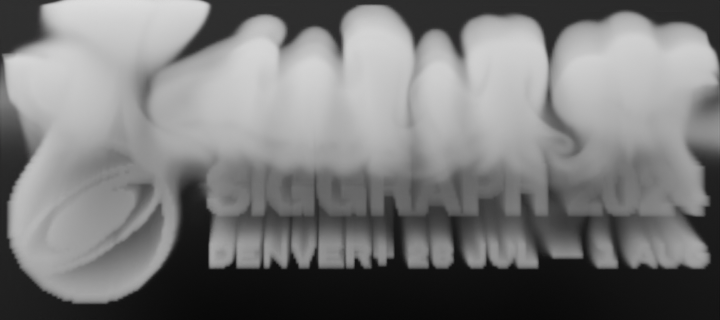}
    \includegraphics[width=0.16\linewidth]{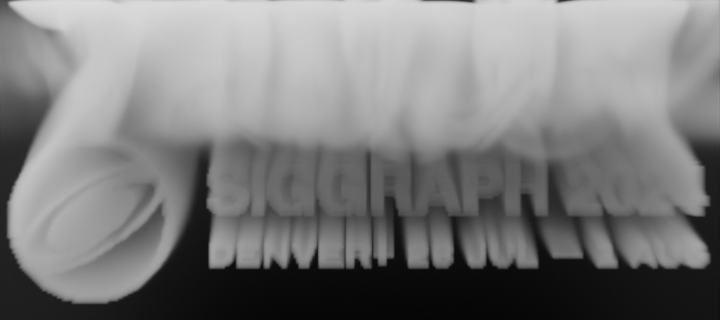}\\
    \caption{Buoyancy simulation in 3D. We simulate a smoke plume rising toward a bunny-shaped obstacle (top, resolution $128^3$) and a sphere obstacle (middle, resolution $128^3$) in 3D using the Boussinesq buoyancy model. We can also incorporate a more complicated smoke source shape as well (bottom, resolution $288\times128\times128$). We rendered the images with the principled volume shader in Blender~\cite{blender2023}.\hfill\videotime{0}{15}}
    \label{fig:3dbuoyancy} 
\end{figure*}

\paragraph{Buoyancy}
In addition to the velocity itself, we can additionally simulate the advection and diffusion of the temperature field $T$ and the smoke concentration field $s$ in a manner similar to the velocity field to add buoyancy effects (Figures \ref{fig:teaser}, \ref{fig:buoyancy}, and \ref{fig:3dbuoyancy}). We use the Boussinesq approximation, which assumes the fluid density variation to be negligible, so that the buoyancy force can be computed using %
\mbox{$\acc = [\alpha \concentration - \beta (\temperature - \temperature_\text{ambient})] \gravity,$}
where $\gravity$ is the gravitational constant vector, $\alpha$ and $\beta$ are positive parameters, and $\temperature_\text{ambient}$ is the ambient temperature constant \cite{Foster1997, Fedkiw2001}.
Naively incorporating buoyancy into the vorticity-based Monte Carlo formulation by \mycitet{RiouxLavoie2022:Fluids} would require the curl of $\acc$~\cite{Park2005}, which would, in turn, require finite difference approximations with some additional errors and ruin the cache-structure-agnostic nature of pointwise estimators.

\paragraph{Divergence control}
As our formulation is based on the standard pressure solve framework, we can easily add velocity sinks and sources in our simulation for better artistic control (\cref{fig:buoyancy} (b)) or other expansion/contraction effects, by modifying the pressure solve step similar to the grid-based formulation~\cite{Feldman2003:Animating}.
We add another term to the right-hand side of \cref{eq:pressure_poisson}, which leads to an additional integral term in our formulation.
Our implementation supports Dirac delta-type sinks and sources, and we sample the location of sinks and sources directly. It should also generalize straightforwardly to more general volumetric sinks and sources with an appropriate sampling technique. This divergence control technique is yet another advantage of adopting a velocity-based formulation, as it is impossible under vorticity formulations.

\paragraph{Viscous fluids}
We test our solver in the case of viscous fluids using the von K\'{a}rm\'{a}n vortex street scenario (\cref{fig:vortexstreet} (e) to (g)). For this simulation, as we increase the viscosity coefficient, or equivalently, as we decrease the Reynolds number, the flow is expected to become less turbulent. We observe that our simulation qualitatively produces the expected type of flow for each Reynolds number configuration we tested. For all the results other than those in \cref{fig:vortexstreet} (e) to (g), we disabled the diffusion step. 

\paragraph{Advection-reflection solver}
In contrast to the case of high viscosity, we may want to introduce less artificial viscosity to the solver. It is known that even without the diffusion step, grid-based solvers suffer from artificial viscosity in the solver because of the dissipation of energy in the advection and projection steps. To address this problem, \citet{Zehnder2018} recently introduced an advection-reflection solver, which still uses the same building blocks of advection and projection steps as the standard advection-projection solvers use, but in a more careful way, to reduce artificial viscosity with the same number of projection steps. We adapted its second-order variant~\cite{Narain2019} to our Monte Carlo solver (\cref{fig:vortexstreet} (h)), and we observe our method similarly benefits from the use of the advection-reflection method, producing more turbulent animation.

\paragraph{Monte Carlo PIC/FLIP}
The numerical diffusion effect we have discussed above is partially due to the semi-Lagrangian advection scheme with some grid data interpolation that our method and many traditional methods employ.
To address this problem in the context of classical grid-based solvers, the PIC/FLIP methods~\cite{zhu2005animating, Harlow1962:PIC, BRACKBILL1986:FLIP} have been popularized.
The idea of PIC/FLIP is that we handle the advection of velocity using particles by (forward) Lagrangian advection, transfer the velocity from particles to grid, perform projection on the grid, and update particle velocities by transferring grid data back to the particles. PIC and FLIP differ in their last grid-to-particle transfer step: PIC transfers the velocity itself whereas FLIP transfers the velocity update. FLIP is expected to produce more turbulent and noisy results. One can also blend the two transfer methods. Our velocity-based Monte Carlo framework can utilize this idea straightforwardly (\cref{fig:vortexstreet} (i) to (k)). In our version, at each step, we first perform the standard particle-to-grid transfer. Then, instead of advecting the particles using the projected grid velocities at the particles' locations via interpolation, we evaluate the velocity (or velocity update) from the grid data exactly at the positions required for particle advection. Exploiting the pointwise estimation capability in this way lets us avoid extra interpolation errors for this step.\looseness=-1

\paragraph{Performance}
We measured the timing of our simulations in \cref{fig:vortexstreet} using their GPU implementations against a basic single-thread CPU implementation of traditional grid-based fluids on a workstation with an Intel Xeon Silver 4316 CPU and an NVIDIA RTX A5000 GPU. Stable fluids with cut-cell boundary handling~\cite{Batty2007} (a) runs at 0.064 seconds per time step, while the fastest implementation of our solver that caches the velocity field after each substep (b) runs at 7.8 seconds per time step, two orders of magnitude slower. If we disable the VPL with this setup, the runtime increases to 109 seconds per time step. For the other caching choices, the runtime for the one without caching after projection (c) increases to 32.1 seconds per time step because we perform the projection at all points used in the RK3 advection, and the one without caching after advection (d) runs at 10.1 seconds per time step due to the increased number of evaluation points for the advection. In general, the walk-on-boundary method for handling obstacles requires a large number of paths, and further efficiency improvements would be desirable to improve the method's practicality. The addition of diffusion steps in (e) to (g) resulted in the increase of runtime by 48\% (e) to 84\% (g) compared to their baseline counterpart implementation (c).

\begin{figure*}
    \centering
    \begin{minipage}[b]{0.138\linewidth}
        \includegraphics[width=\linewidth]{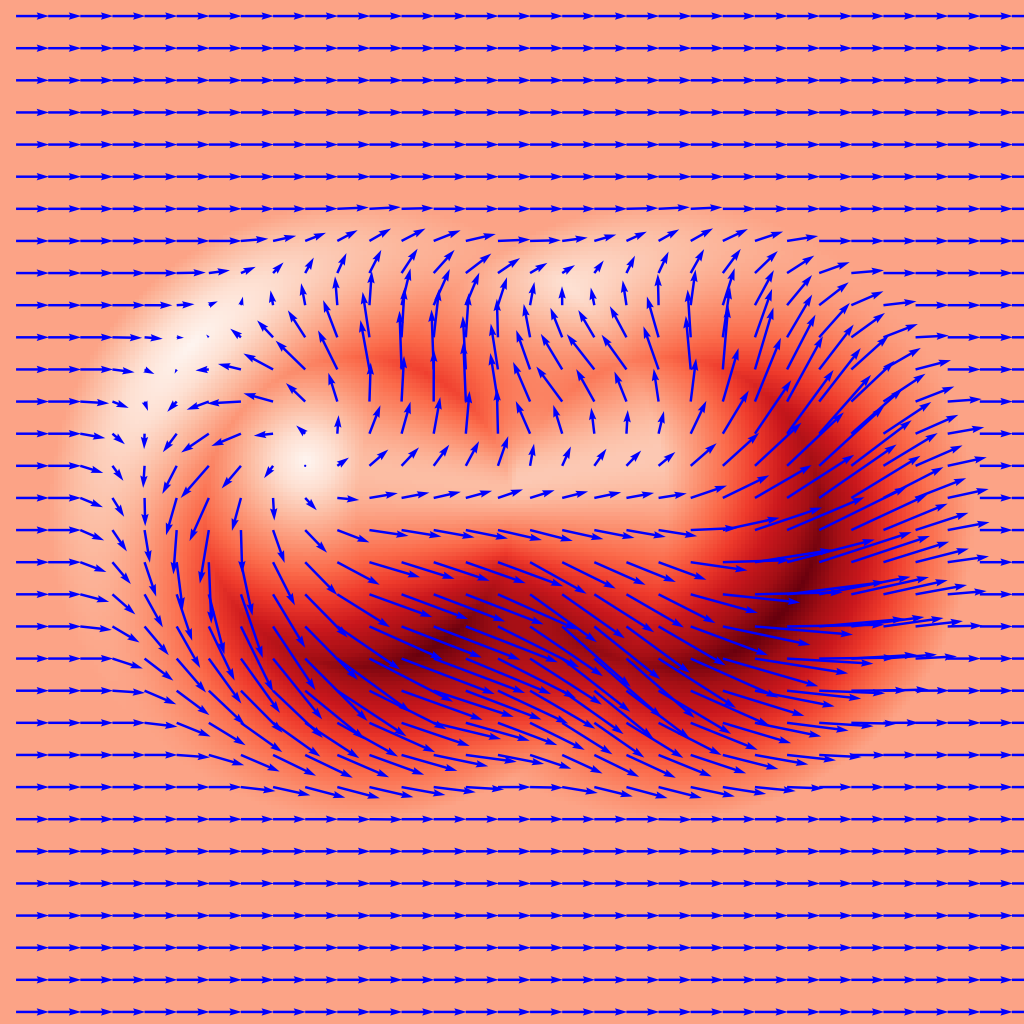}\\
        \centering\textsf{{\small Before Projection}}\\
        \vspace{0.7em}
        \includegraphics[width=\linewidth]{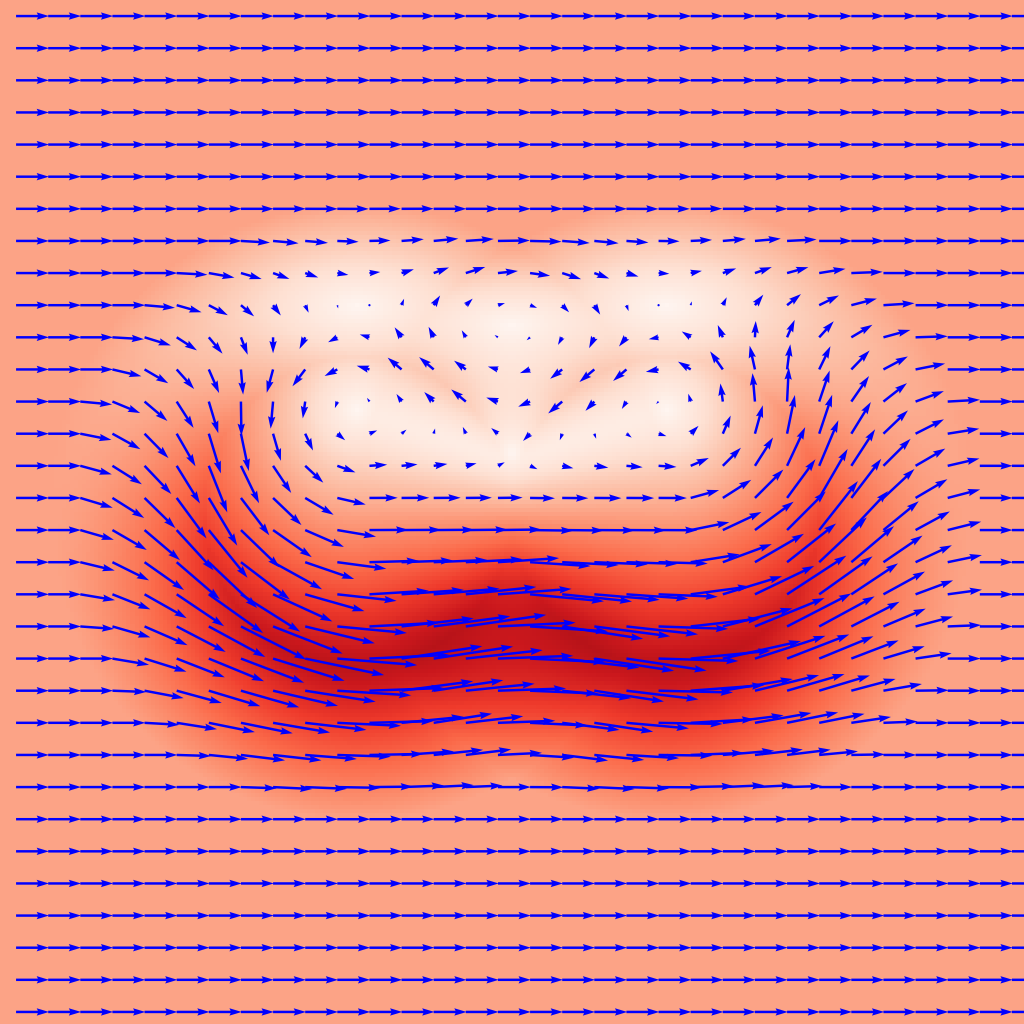}\\
        \centering\textsf{{\small After Projection}}\\
        \vspace{0.3em}
    \end{minipage}
    \begin{minipage}[b]{0.02\linewidth}
    \end{minipage}
    \includegraphics{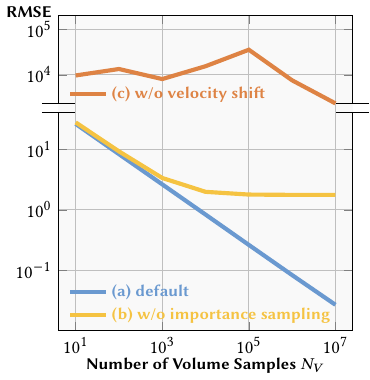}
    \includegraphics{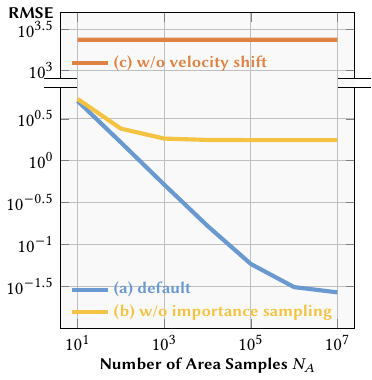}
    \caption{Projection error without solid boundaries. Our projection step projects out the divergent velocity mode from the velocity field, converting the top left field into the one on the bottom left (resolution $256^2$). We visualize the velocity field using arrows and we show its magnitude using the darkness of the red color as well.
We measure the root mean squared errors measured against the analytical solution and plot them by altering the number of volume samples $N_V$ with a fixed number of area samples $N_A=10^7$ in the middle and by altering $N_A$ with $N_V=10^7$ on the right. We observe the inverse square root convergence rate in both cases with our formulation with the default sampling strategy (a) as described in \cref{sec:projection_wo_boundaries}, while the bias due to the volume term eventually dominates the error on the plot for $N_A$ as we increase $N_A$. We also tested the estimator by (separately) replacing the volume term importance sampling according to the PDF proportional to $1/r^{(d-1)}$ with a uniform sampling (b), 
    and disabling the global velocity shift described in \cref{eq:pressure_grad_final} (c).
    Both of these alternatives either converge slower than the proposed method or fail to converge.}
    \label{fig:convergence}
\end{figure*}

\paragraph{Error analysis}
We performed a convergence test for the projection step in the absence of obstacles. The estimator is unbounded in this case, and we can observe in \cref{fig:convergence} that the error decreases at the inverse square root rate, as we increase the number of samples to estimate the volume term and the area term, respectively. We also observe that using a simpler formulation or a simpler sampling technique makes the estimator converge slower or completely fail to converge. In \cref{fig:2dbunny}, we compared our simulations with different numbers of samples. We observe that using low sample counts can blow up the simulation, and the noise of the velocity estimate is typically larger around the boundary. As \citet{Sugimoto2023} discuss, the variance of the walk-on-boundary method is particularly large for interior Neumann problems in non-convex domains, and \cref{fig:2dbunny} shows such a scene.

\section{Conclusion and Discussion}\label{sec:discussion}
We have presented a velocity-based Monte Carlo method for fluid simulation. We developed a pointwise estimator for each substep and demonstrated flexible integration of existing velocity-based techniques into our method.

One could also consider using our velocity-based pointwise estimators to design a cacheless scheme without any spatial discretization whatsoever, similar to the vorticity-based strategy described by \mycitet{RiouxLavoie2022:Fluids}. 
As \mycitet{RiouxLavoie2022:Fluids} pointed out, this kind of fully recursive strategy offers truly pointwise estimation at the cost of large variance and thus has a high computational cost. While of theoretical value, this approach is impractical today because the estimation cost grows exponentially as the simulation proceeds in time, depending on the specific path sampling strategy.

Our method's computational speed is not yet comparable to traditional methods as a relatively large number of samples is required to produce results with low enough error. Moreover, having boundaries in the scene necessitates the use of an advanced Monte Carlo solver for (Neumann) boundary value problems, with corresponding increased variance, leading to higher costs.

In addition to efficiency improvements, our method will benefit from any future improvements to the walk-on-boundary method. Support for spatially varying diffusion equations under the walk-on-boundary method, analogous to what was proposed for the walk-on-spheres method~\mycite{Sawhney:2022:GFMC}, would allow us to support variable fluid densities in our framework.
Applying differentiable rendering techniques similar to \citet{Yilmazer2022} would enable the use of our solver for inverse problems.
Supporting double-sided boundary conditions by extending the boundary integral formulation would also be a useful extension.

We designed our solver with a velocity-only formulation, but it may be beneficial to revisit this choice as (grid-based) methods utilizing velocity gradient information \cite{Jiang2015:APIC, Nabizadeh2022:Covector, Feng2023:Impulse} have recently shown promising improvements. Other ideas in the velocity-based fluid literature, such as the method of characteristic mapping \cite{Qu2019:Efficient} to lessen advection errors or energy-preserving integrators \cite{Mullen2009:EnergyPreserving} to reduce splitting errors, are also worth investigating in our Monte Carlo context.

Lastly, we believe our method paves the way for a full Monte Carlo simulation of liquids with free surfaces, and we look forward to the continued advancement of Monte Carlo techniques for physical simulations more broadly.

\begin{acks}
This research was partially funded by NSERC Discovery Grants (RGPIN-2021-02524 \& RGPIN-2020-03918), CFI-JELF (Grant 40132), and a grant from Autodesk. 
Ryusuke Sugimoto was partially funded by David R. Cheriton Graduate Scholarship. 
This research was enabled in part by support provided by SHARCNET and the Digital Research Alliance of Canada.
We thank Joel Wretborn, Brooke Dolny, Rikin Gurditta, and Clara Kim for proofreading.
We would like to thank the anonymous reviewers for their constructive evaluations and feedback.
\end{acks}

\newpage

\bibliographystyle{ACM-Reference-Format}
\bibliography{main}

\definecolor{mycolor1}{HTML}{6A99D0}
\definecolor{mycolor2}{HTML}{DE8344}
\definecolor{mycolor3}{HTML}{7EAB55}
\definecolor{mycolor4}{HTML}{F5C342}

\begin{figure*}
    \begin{minipage}[b]{0.015\linewidth} \centering
    \begin{sideways}\textsf{\small\;$\times10$ samples}\end{sideways}
    \end{minipage}
    \includegraphics[trim={0.13cm 0cm 0.13cm 0cm}, clip, width=0.135\linewidth]{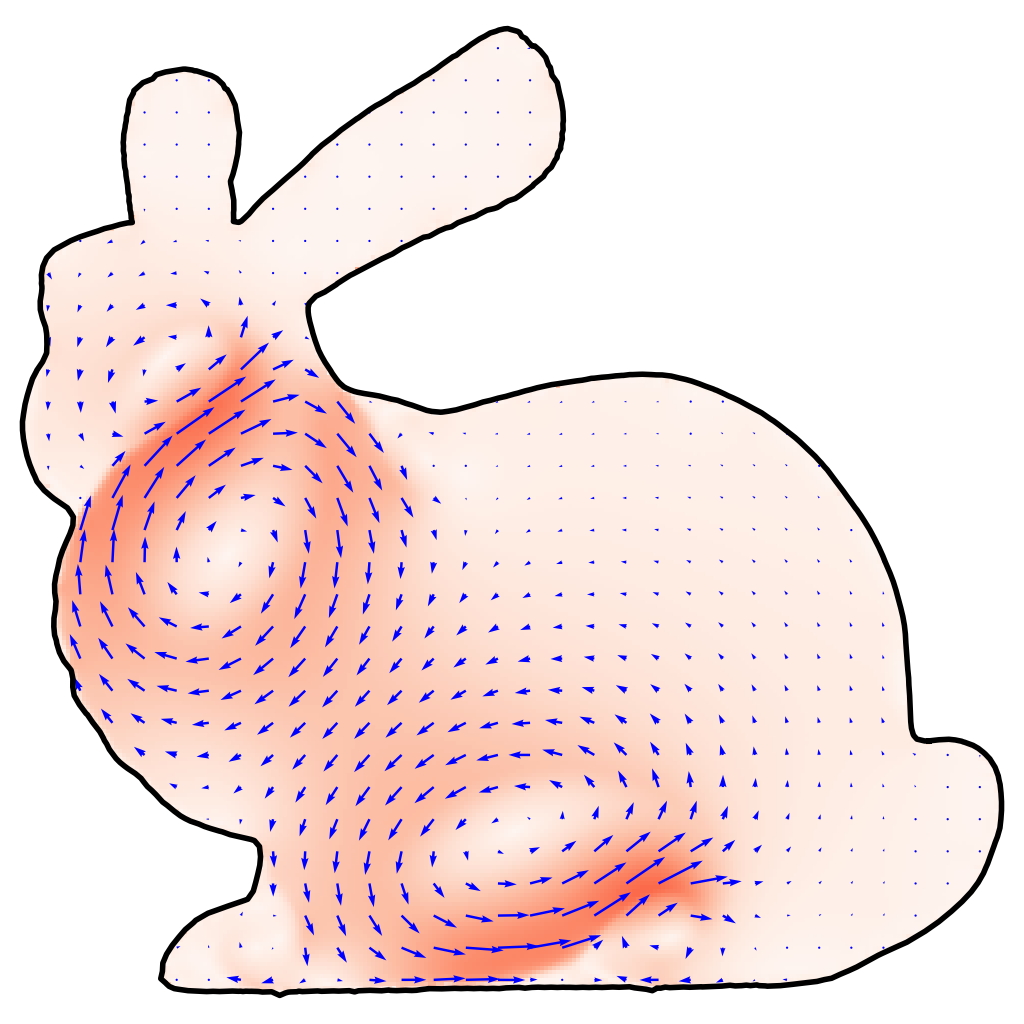}
    \includegraphics[trim={0.13cm 0cm 0.13cm 0cm}, clip, width=0.135\linewidth]{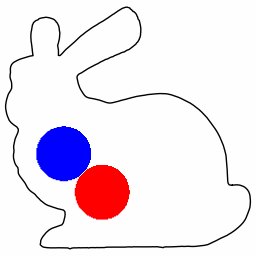}
    \includegraphics[trim={0.13cm 0cm 0.13cm 0cm}, clip, width=0.135\linewidth]{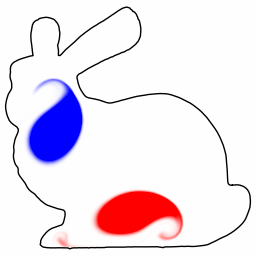}
    \includegraphics[trim={0.13cm 0cm 0.13cm 0cm}, clip, width=0.135\linewidth]{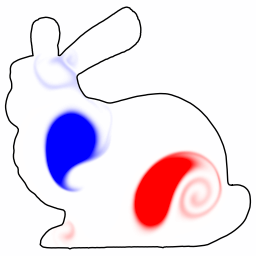}
    \includegraphics[trim={0.13cm 0cm 0.13cm 0cm}, clip, width=0.135\linewidth]{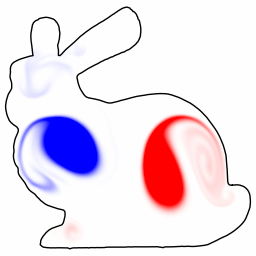}
    \includegraphics[trim={0.13cm 0cm 0.13cm 0cm}, clip, width=0.135\linewidth]{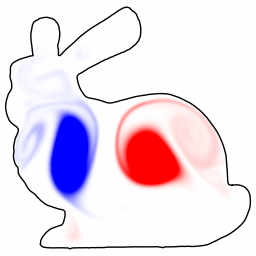}
    \includegraphics[trim={0.13cm 0cm 0.13cm 0cm}, clip, width=0.135\linewidth]{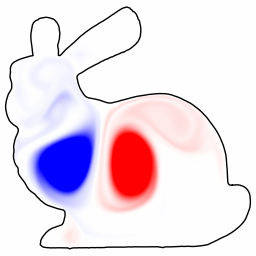}\\

    \begin{minipage}[b]{0.015\linewidth} \centering
    \begin{sideways}\textsf{\small\;$\times1$ samples}\end{sideways}
    \end{minipage}
    \includegraphics[trim={0.13cm 0cm 0.13cm 0cm}, clip, width=0.135\linewidth]{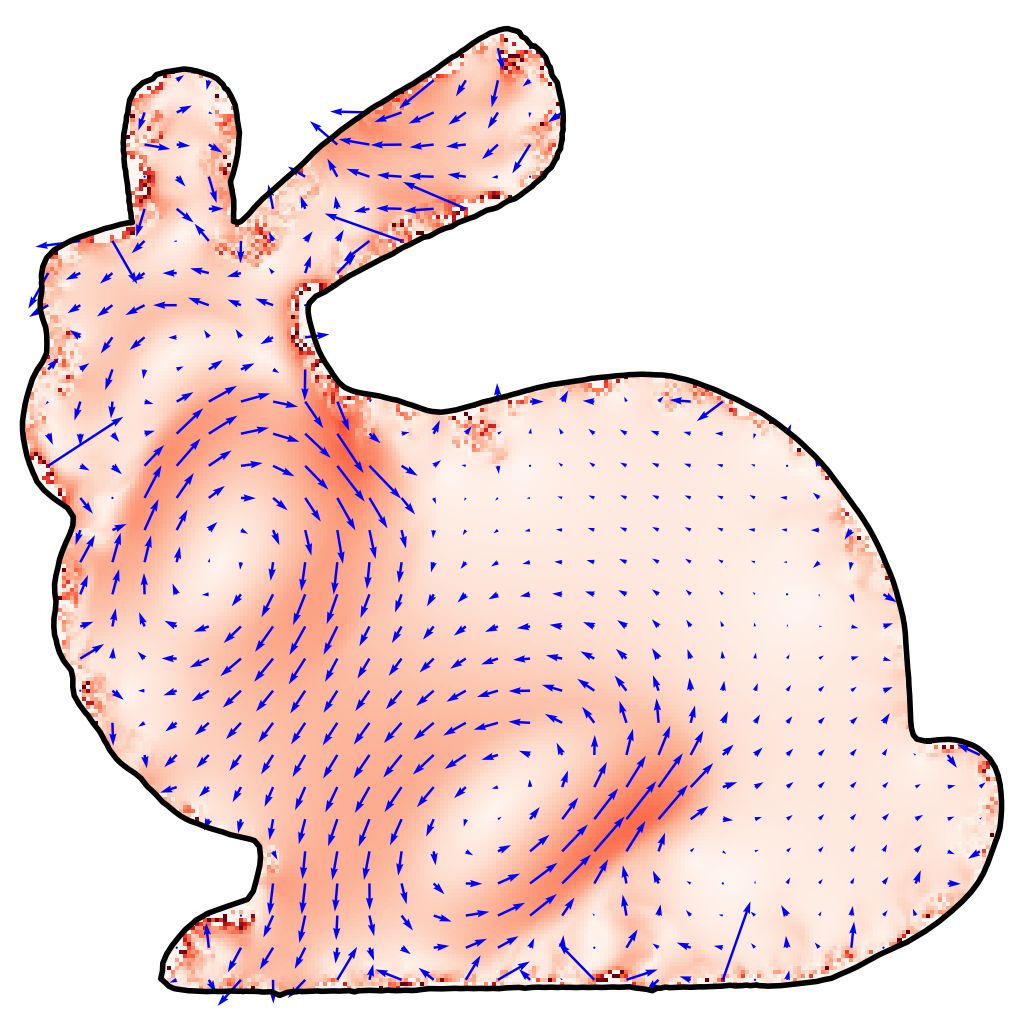}
    \includegraphics[trim={0.13cm 0cm 0.13cm 0cm}, clip, width=0.135\linewidth]{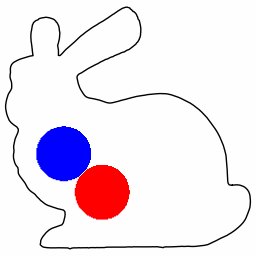}
    \includegraphics[trim={0.13cm 0cm 0.13cm 0cm}, clip, width=0.135\linewidth]{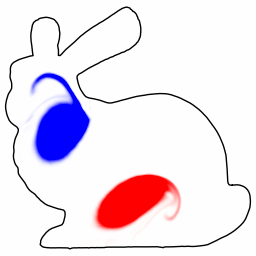}
    \includegraphics[trim={0.13cm 0cm 0.13cm 0cm}, clip, width=0.135\linewidth]{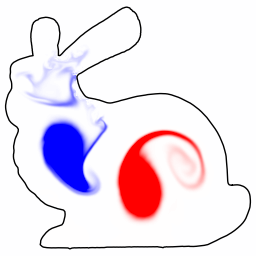}
    \includegraphics[trim={0.13cm 0cm 0.13cm 0cm}, clip, width=0.135\linewidth]{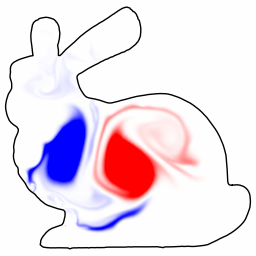}
    \includegraphics[trim={0.13cm 0cm 0.13cm 0cm}, clip, width=0.135\linewidth]{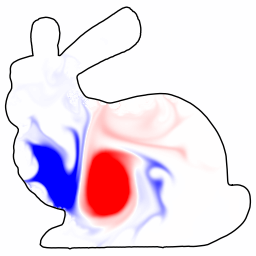}
    \includegraphics[trim={0.13cm 0cm 0.13cm 0cm}, clip, width=0.135\linewidth]{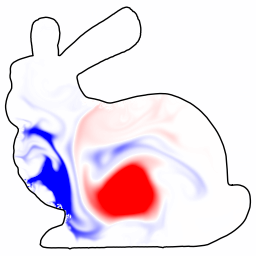}\\

    \begin{minipage}[b]{0.015\linewidth} \centering
    \begin{sideways}\textsf{\small\;$\times0.1$ samples}\end{sideways}
    \end{minipage}
    \includegraphics[trim={0.13cm 0cm 0.13cm 0cm}, clip, width=0.135\linewidth]{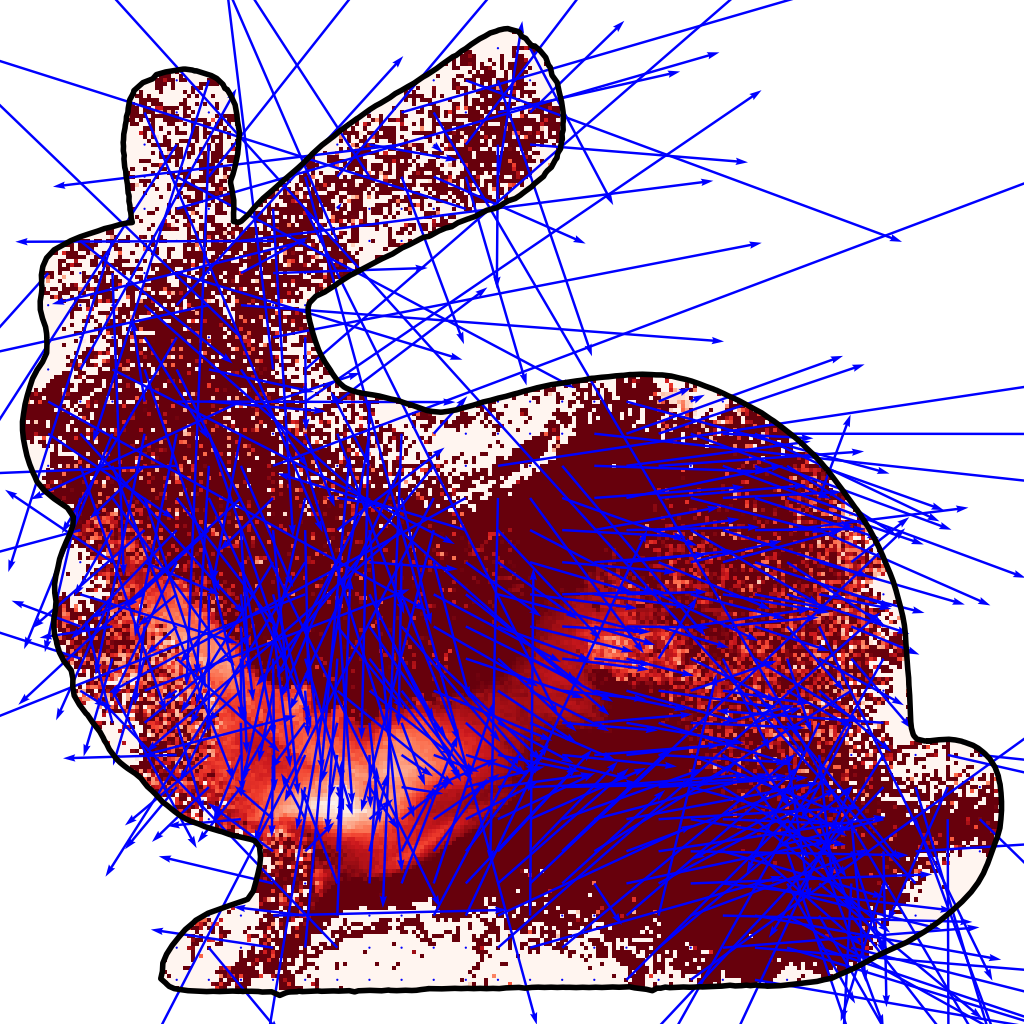}
    \includegraphics[trim={0.13cm 0cm 0.13cm 0cm}, clip, width=0.135\linewidth]{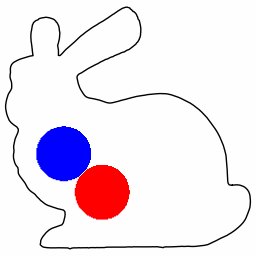}
    \includegraphics[trim={0.13cm 0cm 0.13cm 0cm}, clip, width=0.135\linewidth]{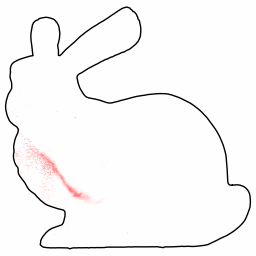}
    \includegraphics[trim={0.13cm 0cm 0.13cm 0cm}, clip, width=0.135\linewidth]{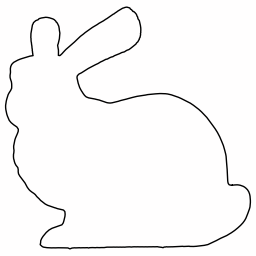}
    \includegraphics[trim={0.13cm 0cm 0.13cm 0cm}, clip, width=0.135\linewidth]{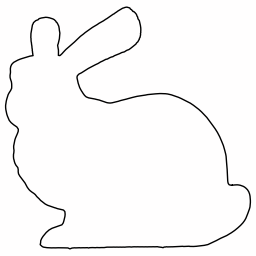}
    \includegraphics[trim={0.13cm 0cm 0.13cm 0cm}, clip, width=0.135\linewidth]{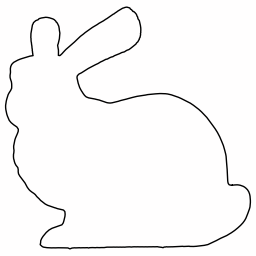}
    \includegraphics[trim={0.13cm 0cm 0.13cm 0cm}, clip, width=0.135\linewidth]{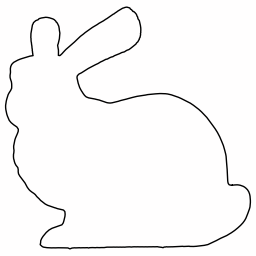}\\

    \begin{minipage}[t]{0.015\linewidth}\;
    \end{minipage}
    \begin{minipage}[t]{0.135\linewidth}\centering\textsf{{\small Velocity Field\\at Frame 40}}\end{minipage}
    \begin{minipage}[t]{0.135\linewidth}\centering\textsf{{\small Frame 0}}\end{minipage}
    \begin{minipage}[t]{0.135\linewidth}\centering\textsf{{\small Frame 40}}\end{minipage}
    \begin{minipage}[t]{0.135\linewidth}\centering\textsf{{\small Frame 80}}\end{minipage}
    \begin{minipage}[t]{0.135\linewidth}\centering\textsf{{\small Frame 120}}\end{minipage}
    \begin{minipage}[t]{0.135\linewidth}\centering\textsf{{\small Frame 160}}\end{minipage}
    \begin{minipage}[t]{0.135\linewidth}\centering\textsf{{\small Frame 200}}\end{minipage}

    \caption{Flow in a 2D bounded non-convex domain (resolution $256^2$). We simulate colored smoke advected by a velocity field induced by a vortex pair inside a bunny shape. On the left, the velocity field at the 40th frame is visualized. In these simulations, the number of samples used in each simulation is 10 times (top), 1 times (middle), and 0.1 times (bottom) the number of samples used in other 2D scenes in this paper. We observe that  using a small sample count can lead to outright failure of the simulation (bottom). Using a modest amount of samples (middle) in this scenario still exhibits large errors and jitter in the animation (see the supplemental video), likely due to larger velocity errors near the boundary and large globally correlated error from the boundary value caching strategy. With the highest sample count (top) the simulation yields a smooth velocity field.\hfill\videotime{4}{18}}
    \label{fig:2dbunny} 
\end{figure*} 

\newpage

\appendix

\newcommand{\given}{^*}

\section{walk-on-boundary for projection}
For \cref{eq:wob_pressure_grad_vel_only} and \cref{eq:wob_density_grad_vel_only}, we defined the normal directions as pointing outward from the fluid domain to describe both interior and exterior problems with a single pair of equations, whereas the previous work  \cite{SabelfeldSimonov1994, Sugimoto2023} had two separate pairs of equations for interior and exterior problems. Hence, our equations for exterior problems differ from \citet{SabelfeldSimonov1994} and \citet{Sugimoto2023} in their signs.

\paragraph{Monte Carlo Estimation}
Based on \cref{eq:wob_pressure_grad_vel_only} and \cref{eq:wob_density_grad_vel_only}, we use a walk-on-boundary Monte Carlo estimator~\cite{SabelfeldSimonov1994, Sugimoto2023}.
Note that \cref{eq:wob_density_grad_vel_only} contains the unknown density function $\wobdens$ itself on the left-hand side of the equation and in one of the integrands on the right-hand side, making it a recursive integral equation similar to the rendering equation~\cite{Kajiya:1986:Rendering} used in path tracing for light transport simulation rather than a simple non-recursive one as in \cref{sec:projection_wo_boundaries}. Following \citet{Sugimoto2023}, we truncate the recursion after $M$ steps and multiply the contribution of the longest path by $0.5$ to design a biased estimator.

We can consider various sampling strategies to estimate the solution to the truncated recursive integral equation with Monte Carlo methods. Following \citet{Sugimoto2023}, we use a forward estimator to solve this in our implementation: to sample a path consisting of multiple boundary points, we randomly sample a boundary point first, take a random walk on randomly sampled boundary points, and finally connect them to each individual evaluation point. 
In the below, for $\pdf_U(\vecx_0)$, we use uniform sampling to sample a point $\vecx_0$ on the solid boundary $\boundary$. Then, for $\pdf_R(\vecx_{m+1}|\vecx_m)$, we uniformly randomly sample a direction from a unit hemisphere for a line that goes through the point $\vecx_m$ and uniformly randomly sample one of the intersection points between the line and the solid boundary $\boundary$ to sample the next point $\vecx_{m+1}$. Recursively applying this sampling strategy lets us sample a series of points $\vecx_1, \vecx_2, ..., \vecx_{M+1}$ to form a path, given previous points in the path. We use this strategy because the PDF associated with this specific forward sampling strategy is known to be proportional to the integral kernel~$\dGdnx$ we have in \cref{eq:wob_density_grad_vel_only}.

Compared to the simplest formulation for the Laplace equation~\cite{Sugimoto2023}, we have some additional non-recursive terms in \cref{eq:wob_pressure_grad_vel_only} and \cref{eq:wob_density_grad_vel_only} as a result of the transformations we discussed in the main paper, and we need to consider how to sample such terms, too. We choose to sample the non-recursive contributions on $\partial \domain$ using the forward estimator as well for the purpose of importance sampling while estimating the other terms similarly to \cref{sec:projection_wo_boundaries}. To do so, we first define two estimators for length $m$ contributions, $\estimator{\mu^1_m(\vecx)}$ and $\estimator{\mu^2_m(\vecx)}$, recursively, based on \cref{eq:wob_density_grad_vel_only}: we define the base case for length $1$ contributions as
\begin{equation}\label{eq:mu11_estimator}\estimator{\mu^1_1(\vecx_1)} = \frac{2\dGdnx(\vecx_1, \vecx_0)}{\pdf_R(\vecx_1|\vecx_0)}\,\vecn(\vecx_0)\cdot\left\{\vel_3(\vecx_0)-\vel_3(\vecx_1)\right\},
\end{equation}
\begin{equation}\label{eq:mu12_estimator}\estimator{\mu^2_1(\vecx_0)} = 2\vecn(\vecx_0) \cdot \left\{ -\estimator{E_V(\vecx_0)} - \estimator{E_A(\vecx_0)}+ \vel_3(\vecx_0) - \velsol(\vecx_0)\right\},
\end{equation}
where $\estimator{\mu^1_1(\vecx_1)}$ corresponds to the non-recursive contribution from the first integral in \cref{eq:wob_density_grad_vel_only} and $\estimator{\mu^1_2(\vecx_0)}$corresponds to the rest of the non-recursive contributions in \cref{eq:wob_density_grad_vel_only}.
Then, we define the contributions from longer paths as
\begin{align}
    \estimator{\mu^1_{m+1}(\vecx_{m+1})} &= \frac{-2\dGdnx(\vecx_{m+1}, \vecx_{m})}{\pdf_R(\vecx_{m+1}|\vecx_{m})} \estimator{\mu^1_m(\vecx_m)},\\
    \estimator{\mu^2_{m+1}(\vecx_m)} &= \frac{-2\dGdnx(\vecx_m, \vecx_{m-1})}{\pdf_R(\vecx_{m}|\vecx_{m-1})} \estimator{\mu^2_m(\vecx_{m-1})}.
\end{align}
Note in particular, when we use the above-mentioned uniform line intersection sampling for $\pdf_R(\vecy|\vecx)$, 
\begin{equation}
 \frac{2\dGdnx(\vecx, \vecy)}{\pdf_R(\vecx|\vecy)} = \kappa(\vecy, \vecx) \cdot \text{sgn}(\vecn(\vecx) \cdot(\vecx - \vecy)),
\end{equation}
where $\kappa(\vecy, \vecx)$ is the number of intersection points that the line that goes through the two points has, excluding point $\vecy$, and sgn is the sign function.
Then, based on \cref{eq:wob_pressure_grad_vel_only}, we can estimate the pressure gradient by taking $N_P$ sample paths in addition to the terms in~\cref{eq:mc_pressure_grad}:
\begin{equation}\label{eq:mc_wob_pressure_grad}
\begin{aligned}
&\estimator{\grad_\vecx\ppres(\vecx)}\\
=\,&\estimator{E_V(\vecx)} + \estimator{E_A(\vecx)} \\
&+  \frac{1}{N_P}\sum_{k=1}^{N_P} \left[  -\frac{\grad_\vecx\fund(\vecx, \vecx_0^k)}{2\pdf_U(\vecx_0^k)}  \vecn(\vecx_0^k)\cdot \left\{\vel_3(\vecx_0^k) - \vel_3(\vecx)\right\}\right. \\
&+ \frac{\grad_\vecx\fund(\vecx, \vecx_{M+1}^k)}{2\pdf_U(\vecx_0^k)}\estimator{\mu^1_{M}(\vecx_{M+1}^k)} +  \frac{\grad_\vecx\fund(\vecx, \vecx_M^k)}{2\pdf_U(\vecx_0^k)}\estimator{\mu^2_{M}(\vecx_M^k)}\\
&+ \left.\sum_{m=1}^{M-1} \frac{\grad_\vecx\fund(\vecx, \vecx_{m+1}^k)}{\pdf_U(\vecx_0^k)}\estimator{\mu^1_{m}(\vecx_{m+1}^k} + \frac{\grad_\vecx\fund(\vecx, \vecx_m^k)}{\pdf_U(\vecx_0^k)} \estimator{\mu^2_{m}(\vecx_m^k)}\right].
\end{aligned}
\end{equation}
On the right-hand side, the first line estimates the last two integrals in \cref{eq:wob_pressure_grad_vel_only}, the second line estimates the non-recursive terms in the first integral, the third line estimates the longest path contributions, and the last line estimates the shorter path contributions. Note that while we do not explicitly indicate so, the contributions from paths of any length $\estimator{\mu^1_m(\vecx_{m+1}^k)}$ and $\estimator{\mu^2_m(\vecx_m^k)}$ implicitly depend on the sampled boundary point $\vecx_0^k$ due to the sampling strategy we employ.

To use the estimator \cref{eq:mc_wob_pressure_grad}, the most naive approach would be to generate sample paths for each individual evaluation point, which has a high computational cost. Instead, we use a boundary value caching strategy similar to the virtual point light method in rendering~\cite{keller1997instant}. We first compute the contributions of $N_P$ subpaths up to the point before we connect them to the evaluation points and cache them at $M+1$ boundary points per path. We then connect all of them to all evaluation points. This significantly increases the effective number of sample paths per evaluation point without increasing the computational cost too much while introducing a correlation of estimates between evaluation points. Even this correlation can be preferable for our application because it guarantees that the contributions from these paths changes smoothly across evaluation points. 

\citet[Figure 9]{Sugimoto2023} apply a similar caching technique to their Neumann problem walk-on-boundary solver based on the same single-layer boundary integral formulation as well, but they use a backward estimator, which works only with a less efficient resampled importance sampling strategy. In contrast, our method can utilize a forward estimator with more efficient line intersection importance sampling. This is at the cost of increasing storage per cache point by a small constant multiplication factor and increasing computation per evaluation point when we sum up the contributions from all cache points.

Similar to \cref{sec:projection_wo_boundaries}, for the estimation of volume integrals $\estimator{E_V(\vecx)}$ in \cref{eq:mu12_estimator} and \cref{eq:mc_wob_pressure_grad}, we use the importance sampling strategy so PDF $\pdf_V$ is proportional to $1/r^{(d-1)}$, and let $\mathbf{S}(\vecx, \vecy) = \mathbf{0}$ for all $\vecy$ outside the simulation domain. Additionally, we need to set $\mathbf{S}(\vecx, \vecy) = \mathbf{0}$ for all $\vecy$ inside of solid obstacles, too. We perform this insidedness test by casting a ray from point $\vecy$ in a random direction and taking a sum of its intersection signs. This strategy can be considered a Monte Carlo estimator for the generalized winding number~\cite{Jacobson2013}. We use uniform sampling for $\estimator{E_A(\vecx)}$.

For all 2D examples in the paper except \cref{fig:2dbunny}, we use $N_V = N_A = 5\cdot10^5$ for direct contributions to each evaluation point. For indirect (path) contributions, we use $N_P = 5\cdot10^5$ paths with path length $4$. For each path, the volume and area term sample counts for the pseudo boundary are $N_V=N_A=10$.

\section{walk-on-boundary for diffusion}
To get scalar diffusion equations from \cref{eq:originaldiffusioneq}, we first absorb the viscosity coefficient $\nu$ into the variable to define $\vec\diffusionsol(\vecx, s)=\visc\overline\vel(\vecx, s/\visc)$ and redefine the range of time $s$ accordingly:
\begin{equation}\label{eq:diffusioneq}
    \begin{aligned}
        \frac{\partial \vec\diffusionsol(\vecx, s)}{\partial s} &= \laplace\vec\diffusionsol(\vecx, s)  & \text{for}&\; \vecx \in \domain, s \in (0, \visc\Delta t)\\
        \vec\diffusionsol(\vecx, s) &= {\vec\diffusionsol}\given(\vecx, s) = \overline\vel(\vecx, s/\nu) & \text{for}&\; \vecx \in \boundary, s \in (0, \visc\Delta t), \text{and}\\
        \vec\diffusionsol(\vecx, 0) &= {\vec\diffusionsol}\given(\vecx, 0) =  \overline\vel(\vecx, 0) & \text{for}&\; \vecx \in \domain,\\
    \end{aligned}
\end{equation}
where the asterisk in $\vec\diffusionsol\given$ indicates that it is a given function.  
This is still a vector-valued equation, but because we only need to deal with simple Dirichlet boundary conditions, we can solve the vector diffusion equation \cref{eq:diffusioneq} component-wise. Thus, we describe the scalar diffusion equation solver for one scalar component of $\mathbf{w}$, $w$ below.

We will summarize the diffusion walk-on-boundary solver from the book by \citet[Chapter 4]{SabelfeldSimonov1994} here, focusing on its use in our context. For more general cases not described here, such as Neumann problems and nonhomogeneous problems, readers should refer to the book. The idea of the diffusion walk-on-boundary method is very similar to the one for the Poisson equation: we convert the partial differential equation into an integral equation and solve it using a ray-tracing-style solver. However, with the additional time dependency, we need to additionally consider the time variation in the diffusion equation and take walks in the space-time domain.
Note that this solver does not discretize the time domain within each time step, unlike most traditional methods that discretize the time with a finite difference.

First, we define the fundamental solution for the diffusion equation \cref{eq:diffusioneq}, also known as the heat kernel,
\begin{equation}
    \diffusionfund(\vecx, s;\, \vecy, \tau) = \heaviside(t') (4\pi t')^{-d/2}e^{-r^2/4t'},
\end{equation}
where $t' = s-\tau$ and $\heaviside(\cdot)$ is the Heaviside step function.

Using the fundamental solution, we can write the solution to the diffusion equation in the form of the double layer potential and an additional initial condition term for $\vecx\in \domain$ and $s\in [0, \nu\Delta t]$:
\begin{equation}\label{eq:diffusion_double_layer}
\begin{aligned}
    \diffusionsol(\vecx, s) =& -\int_0^s \int_\boundary \dZdny(\vecx, s;\, \vecy, \tau) \diffusionwobdens(\vecy, \tau) \dAy \dtau\\
    &+\int_\domain \diffusionfund(\vecx, s; \vecy, 0 ) \diffusionsol\given(\vecx, 0) \dVy,
\end{aligned}
\end{equation}
where $\diffusionwobdens(\cdot, \cdot)$ is an unknown density function defined for $\vecx\in \boundary$ and $s\in [0, \nu\Delta t]$.
We can also derive a recursive boundary integral equation for $\diffusionwobdens(\cdot, \cdot)$ by taking the limit $\vecx\rightarrow\boundary$ in \cref{eq:diffusion_double_layer}:
\begin{equation}\label{eq:diffusion_bie}
\begin{aligned}
    \diffusionwobdens(\vecx, s) =& \int_0^s \int_\boundary 2\dZdny(\vecx, s;\, \vecy, \tau) \diffusionwobdens(\vecy, \tau) \dAy \dtau\\
     &+ 2 \diffusionsol\given(\vecx, s) - 2\int_\domain \diffusionfund(\vecx, s; \vecy, 0 ) \diffusionsol\given(\vecx, 0) \dVy.
\end{aligned}
\end{equation}

\paragraph{Monte Carlo Estimation} We design a Monte Carlo estimator based on \cref{eq:diffusion_double_layer} and \cref{eq:diffusion_bie}.
First, we define an estimator for the initial condition term with $N_I$ samples using PDF $\pdf_I$ as
\begin{equation}\label{eq:EI_diffusion}
\estimator{E_I(\vecx, s)}= \frac{1}{N_I}\sum_{i=1}^{N_I} \frac{\mathbf{Z}(\vecx, s;\vecy^i, 0)}{\pdf_I(\vecy^i|\vecx, s)}\diffusionsol\given(\vecy^i, 0).
\end{equation}
Using this estimator, we can define the estimator for \cref{eq:diffusion_double_layer} with $N_D$ path samples using PDF $\pdf_D$ as
\begin{equation}\label{eq:mc_diffusion_sol}
    \estimator{\diffusionsol(\vecx, s)} = \estimator{E_I(\vecx, s)} +  \frac{1}{N_D}\sum_{j=1}^{N_D}  -\frac{\dZdny(\vecx, s;\, \vecy^j, \tau^j)}{\pdf_D(\vecy^j, \tau^j|\vecx, s)} \estimator{\diffusionwobdens(\vecy^j, \tau^j)}
\end{equation}
and the one for \cref{eq:diffusion_bie} with $1$ sample with PDF $\pdf_E$ as
\begin{equation}\label{eq:mc_diffusion_bie}
\begin{aligned}
    \estimator{\diffusionwobdens(\vecx, s)} =  2\frac{\dZdny(\vecx, s;\, \vecy, \tau)}{\pdf_E(\vecy, \tau|\vecx, s)} \estimator{\diffusionwobdens(\vecy, \tau)}
    &+ 2 \diffusionsol\given(\vecx, s) -2\estimator{E_I(\vecx, s)},
\end{aligned}
\end{equation}
which is a recursive estimator: $\estimator{\diffusionwobdens(\vecy, \tau)}$ included on the right-hand side should be estimated recursively.
We use a backward estimator and start generating paths for this recursive estimator from each evaluation point $(\vecx, \nu\Delta t)$ toward time~$0$. Note that $\pdf_D$ and $\pdf_E$ only need to sample points with time $\tau < s$ because the integral kernel $\dZdny$ is zero for $\tau \geq s$. Intuitively, this is because the solution to the heat equation depends only on previous times.
Using this property, we sample points in the space-time domain so that times for the sampled sequence of points strictly decrease.
We terminate the recursion when the sampled time $\tau$ is negative because the original integral domain does not contain the negative part. While the walk-on-boundary method for the Poisson equation is biased, the walk-on-boundary method for the diffusion equation gives an unbiased solution estimate. 

As for the specific sampling strategy, for the initial condition sampling $\pdf_I(\vecy^i|\vecx, s)$, we sample the points by ${\vecy_i\leftarrow \vecx + \sqrt{t \gamma_{d/2}}\vec{\omega}}$, where $\gamma_{d/2}$ is a sample drawn from the Gamma distribution with shape parameter $d/2$ and scale parameter 1, and $\vec{\omega}$ is a uniformly random direction sampled on a unit sphere. With this sampling strategy, we get
\begin{equation}
\frac{\mathbf{Z}(\vecx, s;\vecy^i, 0)}{\pdf_I(\vecy^i|\vecx, s)} = 1.
\end{equation}Similar to \cref{sec:projection_with_boundaries}, we set $\mathbf{Z}(\vecx, s;\vecy^i, 0) = 0$ for all $\vecy$ inside of solid obstacles in \cref{eq:EI_diffusion}.

For $\pdf_D(\vecy, \tau|\vecx, s)$ and $\pdf_E(\vecy, \tau|\vecx, s)$, we sample $\vecy$ using the uniform line intersection sampling as in \cref{sec:projection_with_boundaries} and sample time $\tau$, given $t$, by $\tau \leftarrow s - \frac{\lVert \vecy - \vecx \rVert}{4\gamma_{d/2}}$ to get
\begin{equation}
\begin{aligned}
 \frac{\dZdny(\vecx, s;\, \vecy, \tau)}{\pdf_D(\vecy, \tau|\vecx, s)}  &= -\kappa(\vecx, \vecy) \cdot \text{sgn}(\vecn(\vecy) \cdot(\vecy - \vecx)), \quad\text{and}\\
 2\frac{\dZdny(\vecx, s;\, \vecy, \tau)}{\pdf_E(\vecy, \tau|\vecx, s)}  &= -\kappa(\vecx, \vecy) \cdot \text{sgn}(\vecn(\vecy) \cdot(\vecy - \vecx)).
\end{aligned}
\end{equation}
In the above, the left-hand side scaling factors of the two expressions differ by $2$ because the PDF $\pdf_D$ and $\pdf_E$ differ even though we use the same strategy: point $\vecx$ for $\pdf_D$ lies within the domain, whereas point $\vecx$ for $\pdf_E$ lies on a boundary. 

\sloppy To draw samples from the Gamma distribution, we use \mbox{$\gamma_1 \leftarrow -\ln(\alpha)$} for 2D, where $\alpha$ is a uniformly random sample in the range $(0, 1)$, and $\gamma_{3/2} \leftarrow \gamma_1 + \xi^2/2$ for 3D, where $\xi$ is a standard normal sample.

For problems without boundaries, we drop the second term in \cref{eq:mc_diffusion_sol}, and the remaining first term estimates the convolution of the input field and a Gaussian function.

In our implementation, we use $N_I=5\cdot10^5$ initial conditions samples at each evaluation point, with $N_D = 5\cdot10^5$ paths with $N_I = 10$ initial condition samples per bounce. Unlike the walk-on-boundary method for the projection step, we do not share the subpaths among evaluation points as we found that the diffusion walk-on-boundary is typically cheaper, and there was not much need to improve its efficiency relative to the projection.
\end{document}